\documentclass[prd,twocolumn,showpacs,floatfix,nofootinbib]{revtex4-1}
\usepackage{hyperref}
\usepackage{latexsym}
\usepackage{graphicx,epsf,epsfig} 
\usepackage{amssymb,amsmath}
\usepackage{bm}
\usepackage[usenames]{color}
\usepackage{pstricks}
\usepackage{rotating}
\usepackage{feynmf}

\usepackage[utf8]{inputenc}

\usepackage{graphicx}
\usepackage{dcolumn}
\usepackage{bm}
\usepackage{color}

\newcommand{\aap}{{Astron.~Astrophys.}}

\newcommand{\pasp}{{Pub.~Astron.~Soc.~Pacific}}

\newcommand{\mnras}{{Mon.~Not.~R.~Astron.~Soc.}}

\newcommand{\jcap}{{Journal of Cosmology and Astroparticle Physics}}

\begin{document}

\title{Modelling Void Abundance in Modified Gravity}

\author{Rodrigo Voivodic$^{1}$}
\email{rodrigo.voivodic@usp.br}
\author{Marcos Lima$^{1}$}
\author{Claudio Llinares$^{2,3}$}
\author{David F. Mota$^{3}$}

\affiliation{
$^{1}$Departamento de F\'{\i}sica Matem\'atica, Instituto de F\'{\i}sica, Universidade de S\~ao Paulo, CP 66318, CEP 05314-970,  S\~ao Paulo, SP, Brazil \\
$^{2}$Institute for Computational Cosmology, Department of Physics, Durham University, Durham DH1  3LE, UK \\
$^{3}$Institute of Theoretical Astrophysics, University of Oslo, PO Box 1029 Blindern, N-0315 Oslo, Norway
}

\date{\today}

\begin{abstract}
 We use a spherical model and an extended excursion set formalism with drifting diffusive barriers to predict the abundance of cosmic voids in the context of general relativity as well as $f(R)$ and symmetron models of modified gravity. We detect spherical voids from a suite of N-body simulations of these gravity theories and compare the measured void abundance to theory predictions. We find that our model correctly describes the abundance of both dark matter and galaxy voids, providing a better fit than previous proposals in the literature based on static barriers. We use the simulation abundance results to fit for the abundance model free parameters as a function of modified gravity parameters, and show that counts of dark matter voids can provide interesting constraints on modified gravity. For galaxy voids, more closely related to optical observations, we find that constraining modified gravity from void abundance alone may be significantly more challenging. In the context of current and upcoming galaxy surveys, the combination of void and halo statistics including their abundances, profiles and correlations should be effective in distinguishing modified gravity models that display different screening mechanisms.      
\end{abstract}

\maketitle

\section{Introduction}

	The large scale structure of the Universe offers a promising means of 
	probing alternative gravity theories \cite{Brax2, deMartino}. Many models of modified gravity can 
	be parameterized by a scalar degree of freedom that propagates 
	an extra force on cosmologically relevant scales. Viable gravity 
	theories must produce a background expansion that is close to that of 
	a Lambda Cold Dark Matter ($\Lambda$CDM) model in order to satisfy current geometry 
	and clustering constraints, and reduce to general relativity (GR) locally in 
	order to satisfy solar system tests. 
	The first feature may be imposed by construction or restriction of the parameter space 
	whereas the latter feature
	relies on a nonlinear screening mechanism operating e.g. on regions of large 
	density or deep potentials \cite{Brax3}. Examples include $f(R)$ models with 
	the chameleon mechanism \cite{Khoury04, Khoury04b, shaw, gan, Gubser, Navarro}, 
        braneworld models which display the Vainshtein mechanism 
        \cite{Vainshtein72, Babichev, Falck}, and the symmetron model with a symmetry breaking 
        of the scalar potential \cite{Hinterbichler10, Hinterbichler11, Hammami, Davis}. 
        Most viable models of cosmic acceleration via modified gravity are nearly indistinguishable 
        at the background level and may be quite degenerate, even when considering linear perturbation
        effects. However, different screening mechanisms operating on nonlinear scales are 
        quite unique features of each model.  It is therefore highly desirable to explore
        observational consequences that help expose these differences, despite the fact 
        that nonlinear physics and baryonic 
        effects must also be known to similar accuracy at these scales.

	Investigating the nonlinear regime of modified gravity models 
	requires N-body simulations \cite{Oyaizu08, Oyaizu08b, Schmidt09,
	Schmidt09b, Schmidt09c,clifton, Khoury09, 
	Li2, Schmidt10, Ferraro11, Zhao11, Li1, Li13, Wyman13, Arnold, Brax4, Candlish, 
	Hagala, Achitouv4, Hammami, Winther, Barreira}, 
	in which one must solve nonlinear 
	equations for the extra scalar field in order to properly account for 
	screening mechanisms. From simulations one may extract the matter power 
	spectrum on linear and non-linear scales \cite{Oyaizu08b, Schmidt09b, Schmidt09c,
	Khoury09, Li11, Wyman13, Taruya} as well as properties of dark matter halos, such as 
	their abundance \cite{Schmidt09, Schmidt09b, bour, Schmidt09c, Li11, 
	Zhao11, Lombriser13, Wyman13}, bias \cite{Schmidt09, Schmidt10, Zhao11, Wyman13} 
	and profiles \cite{Schmidt09, Schmidt09b, Zhao11, Lombriser12}.
	
	From the theoretical perspective, estimating e.g. the  power spectrum 
	in the nonlinear regime is non-trivial even for GR, and more so for 
	modified gravity \cite{Koyama09, Taruya}, as the screening mechanisms must be properly 
	accounted for in the evolution equations \cite{Brax1}. The halo model \cite{Cooray} provides 
	an alternative to study these nonlinearities \cite{Schmidt09, Schmidt10}, but it has its limitations 
	even in standard GR. 
	Moreover it requires accurate knowledge of various halo properties, including 
	abundance, bias and profiles.
	
	In GR the halo mass function may be estimated from the linear 
	power spectrum and spherical collapse within the Press-Schechter \cite{Press} 
	formalism and its extensions \cite{Sheth2, Bond} 
	or from empirical fits to simulations for higher precision \cite{Tinker, Jenkins}.
	However for modified gravity screening mechanisms operate effectively 
	within the most massive halos, and must be properly accounted for \cite{Li11}.  
	In addition, massive clusters have observational complications such as the 
	determination of their mass-observable relation \cite{Lima05}, which must be known to 
	good accuracy in order for us to use cluster abundance for cosmological purposes.  
         These relations may also change in modified gravity \cite{Arnold}.  
	
	Cosmological voids, i.e. regions of low density and shallow potentials, 
	offer yet another interesting observable to investigate modified 
	gravity models \cite{Clampitt}. Screening mechanisms operate weakly within voids, 
	making them potentially more sensitive to modified gravity effects \cite{Pisani}.  
	One of the main issues for using voids is their very definition, which is not 
	unique both theoretically and observationally. Compared to halos, the properties 
	of voids have not been discussed in as much detail, although there have been a number 
	of recent developments on the theory, simulations and observations of voids 
	\cite{Sanchez, Pisani, Massara, Cai2, Wojtak, Pollina, Nadathur1, Nadathur2, Nadathur3}.
	
	Despite ambiguities in their exact definition, it has been observed in simulations 
	that voids are quite spherical \cite{Sheth1}, and therefore it is expected that the spherical 
	expansion model for their abundance must work well (differently from 
	halos, for which spherical collapse alone is not 
	a very good approximation
	 \cite{Achitouv2}). 
	In this work, we use N-body simulations of $\Lambda$CDM as well as
	$f(R)$ and symmetron models of modified gravity in order to identify 
	cosmic voids and study their abundance distribution. In order to interprete 
	the simulation results, we use    
	a spherical model and an extended excursion set formalism 
	with underdense initial conditions to construct the void distribution function.
	Our extended model includes 
	two drifting diffusive barriers in a similar fashion to the work 
	from \cite{Maggiore1, Maggiore2} to describe halo abundance.
	As a result, our model  accounts for the void-in-cloud effect and 
	generalizes models with static barriers \cite{Jennings}.
	
	We start in \S~\ref{sec:perturbation} describing the parametrization of 
	perturbations in $f(R)$ and symmetron gravity as well as the spherical model 
	equations. In \S~\ref{sec:void_dist} we use the excursion set formalism to 
	model void abundance and in \S~\ref{sec:void_sim} we describe the procedure for 
	void identification from simulations.  Importantly, we define spherical voids 
	in simulations with a criterium that is self-consistent with our predictions. 
	In \S~\ref{sec:res} we present 
	our main results, using simulations to fit for the model free parameters and 
	studying constraints on modified gravity from ideal dark matter voids. We also study 
	the possibility of using our model to describe galaxy voids.
	Finally, in \S~\ref{sec:discussion} we discuss our results and conclude.

\section{Perturbations}\label{sec:perturbation}

	The spherical evolution model is usually the first step to investigate 
	the abundance of virialized objects tracing the Universe structure, 
	such as halos, and likewise it is a promising tool for voids. It also offers a
	starting point to study the collapse of non-spherical structures
	\cite{Achitouv1,Achitouv2} and  the parameters required to quantify the 
	abundance of these objects within extended models \cite{Zentner}.

	The large scale structure of the Universe is well characterized by the 
	evolution of dark matter, which interacts only gravitationally and can be approximated 
	by a pressureless perfect fluid. The line element for a perturbed Friedmann-Lema\^itre-Robertson-Walker (FLRW) 
	metric in the Newtonian gauge is given by 
\begin{equation}
ds ^{2} = -a^{2}(1 + 2\Psi)d\tau ^{2} + a^{2}(1 - 2 \Phi)dl^{2}\,,
\label{Pertubed_metric}
\end{equation}	
where $a$ is the scale factor, $\tau$ is the conformal time related to the 
physical time $t$ by $ad\tau = dt$, $dl^2$ is the line element for the spatial metric in  
a homogeneous and isotropic Universe and $\Psi$ and $\Phi$ are the 
gravitational potentials.
	
	For a large class of modified gravity models, the perturbed fluid equations
	in Fourier space are given by \cite{Brax1}
\begin{eqnarray}
\dot{\delta}  &=& -(1+\delta)\theta \,,\\
\dot{\theta} +2H\theta + \frac{1}{3} \theta ^{2}  &=&  k^{2} \Phi\,, \\
-k^{2} \Phi &=& 4 \pi G \mu (k,a) \bar{\rho}_{\rm m}\delta\,, 
\label{eq:perturbations}
\end{eqnarray}
where $\delta = (\rho_{\rm m} - \bar{\rho}_{\rm m})/\bar{\rho}_{\rm m}$ is the matter density contrast, $\theta$
 is the velocity divergence, 
  $H = \dot{a}/a$ is the Hubble parameter and dots denote derivatives with respect to physical time $t$.

The first is the continuity equation, the second the Euler equation and the last is the modified Poisson equation, where modified gravity effects are incorporated within 
the function $\mu (a,k)$. In general this function depends on scale factor $a$ 
as well as physical scale or wave number $k$ in Fourier space. 

	Combining these equations we obtain an evolution equation for spherical perturbations in modified gravity \cite{Pace} given by
\begin{equation}
\delta '' + \left( \frac{3}{a} + \frac{E'}{E} \right) \delta ' - \frac{4}{3}\frac{(\delta ')^{2}}{1+\delta}  = \frac{3}{2} \frac{\Omega _{m}}{a^{5} E^{2}}\mu(k,a) \delta (1+\delta)\,,
\label{delta_eq}
\end{equation} 
where primes denote derivatives with respect to the scale factor $a$, 
$E(a) = H(a)/H_{0}$, $H(a)$ is the Hubble parameter at $a$, 
$H_0$ is the Hubble constant and $\Omega_m$ is the present matter density relative 
to critical. Clearly the growth of perturbations is scale-dependent -- a 
general feature of modified theories of gravity.
	
	The linearized version of Eq.~\eqref{delta_eq}  is given by
\begin{equation}
\delta '' + \left( \frac{3}{a} + \frac{E'}{E} \right) \delta ' = \frac{3}{2} \frac{\Omega _{m}}{a^{5} E^{2}}\mu(k,a) \delta\,,
\label{delta_lin}
\end{equation}
and can be used to determine linear quantities, such as the linear power spectrum. 
Notice that this matter linear equation is valid more generally and does not not require spherical perturbations.

	The function $\mu(k,a)$ above is given by \cite{Brax1}
\begin{equation}
\mu (k,a) = \frac{(1+2\beta ^{2})k^{2} + m^{2}a^{2}}{k^{2} + m^{2}a^{2}}\,,
\label{mu}
\end{equation}
where $\beta$ is the coupling between matter and the fifth force and $m$ is the mass of the scalar field propagating the extra force.

	It is important to stress that the parameterization in Eq.~\eqref{mu} 
	does not fully account for modified gravity perturbative effects, containing only effects of the background 
	and linear perturbations for extra fields related to modified gravity. 
	This is enough for the linearized Eq.~\eqref{delta_lin}, but is 
	only an approximation in Eq.~\eqref{delta_eq}. For instance the parameterization in Eq.~\eqref{mu} 
	does not contain effects from the screening mechanisms, which would turn 
	$\mu$ into a function not only of scale $k$, but of e.g. the 
	local density or gravitational potential.

\subsection{$f(R)$ gravity}

	The action for $f(R)$ gravity is given by
\begin{equation}
S = \int  d^{4}x \sqrt{-g} \left[ \frac{M_{pl}^{2}}{2}R + f(R) \right]+ S_{m}[g_{\mu \nu}, \psi _{i}]  \,,
\end{equation}
where $g_{\mu\nu}$ is the Jordan frame
metric, $g$ is the metric determinant, 
$M_{pl} ^{2} = (8\pi G)^{-1}$, $G$ is Newton's constant, $R$ is the Ricci scalar
and $S_m$ is the action for the matter fields $\psi_i$ minimally coupled to the metric. 
For concreteness, we will employ the parameterization of 
	Hu \& Sawicki \cite{Hu}, which in the large curvature regime can 
	be expanded in powers of $R^{-1}$ as
\begin{equation}
f(R) \approx -16\pi G \rho _{\Lambda} -\frac{f_{R0}}{n}\frac{R_{0}^{n+1}}{R^{n}} \,,
\label{fr}
\end{equation}
where the first constant term is chosen to match a $\Lambda$CDM expansion, 
such that $\rho _{\Lambda}$ is the effective dark energy density (of a cosmological 
constant $\Lambda$ in this case) in the 
late-time Universe, and $f_{R0}$ and $n$ are free parameters.
Here $f_R \equiv df/dR$ represents an extra scalar 
degree of freedom propagating a fifth force, such that $f_{R0}$ denotes the background 
value of this scalar field at $z=0$. We fix $\Lambda$ such that $\Omega_\Lambda=0.733$ 
and $n=1$ to reflect the values used in the simulations to be described in \S~\ref{sec:void_sim}.

	It can be shown that $f(R)$ models are a particular class of scalar-tensor theories, for which 
	the parameters from Eq.~\eqref{mu} are \cite{Brax1}
\begin{eqnarray}
\beta &=& \frac{1}{\sqrt{6}}\,, \nonumber \\
m(a) &=& m_{0}\left( \frac{\Omega _{m} a^{3} + 4\Omega _{\Lambda}}{\Omega _{m} + 4\Omega _{\Lambda}}\right)^{(n+2)/2} \,,
\label{para_fr}
\end{eqnarray}
where
\begin{equation}
m_{0} = \frac{H_{0}}{c}\sqrt{ \frac{\Omega _{m} + 4\Omega _{\Lambda}}{(n+1)f_{R0}}}\,.
\end{equation}

	Solving Eqs.~\eqref{delta_eq} and \eqref{delta_lin} numerically given 
	initial conditions where the Universe evolution was similar to that from GR, 
	it is possible to compute important parameters for characterizing 
	the abundance of cosmic voids.
	
\subsection{Symmetron}

	The symmetron model is described by the action \cite{Davis}
\begin{eqnarray}
S &=& \int d^{4}x \sqrt{-\tilde{g}} \left[ \frac{M_pl^{2}}{2}\tilde{R} - \frac{1}{2}\partial _{\mu} \phi \partial ^{\mu} \phi - V(\phi) \right] \nonumber \\
&+& S_{m}[g_{\mu \nu}, \psi _{i}] \,,
\label{Symmetron}
\end{eqnarray}
where $\phi$ is the symmetron field, $V(\phi)$ is the field potential, $S_{m}[g_{\mu \nu}, \psi _{i}]$ is the action for 
the matter fields $\psi _{i}$ and $\tilde{g}_{\mu \nu}$ is the Einstein frame metric related with the Jordan frame metric via the conformal rescaling

\begin{equation}
g_{\mu \nu} = A^{2}(\phi) \tilde{g}_{\mu \nu} \,,
\end{equation}
and $\tilde{R}$ is the corresponding Einstein frame Ricci scalar.

	The coupling function $A(\phi)$ and the field potential $V(\phi)$ are chosen to be polynomials satisfying the parity symmetry $\phi \rightarrow -\phi$
\begin{eqnarray}
A(\phi) &=& 1 + \frac{1}{2}\left( \frac{\phi}{M} \right) ^{2}\,, \\
V(\phi) &=& V_{0} - \frac{1}{2} \mu ^{2} \phi ^{2} + \frac{1}{4} \lambda \phi ^{4} \,,
\label{Pot/cou}
\end{eqnarray}
where $M$ and $\mu$ have dimensions of mass and $\lambda$ is dimensionless. We assume that $\left(\phi/{M} \right)^{2} \ll 1$, so that the coupling function can indeed be expanded up to second order.

The mass and coupling parameters of the field (see Eq.~\eqref{mu}) are \cite{Davis}
\begin{eqnarray}
m_{\phi} ^{2}(a) &=& \left\{ \begin{array}{ll} 
\mu ^{2} \left( \frac{\bar{ \rho } _{m}(a)}{\rho _{SSB}} - 1 \right), \quad \bar{ \rho} _{m} > \rho_{SSB} \\
2 \mu ^{2} \left( 1 - \frac{\bar{\rho }_{m}(a)}{\rho _{SSB}} \right), \quad \bar{ \rho} _{m} < \rho_{SSB}
\end{array} \right. \nonumber \\
\beta(a) &=& \beta _{0} \frac{\phi (a)}{\phi _{0}} \,,
\label{para_symm}
\end{eqnarray}
where $\rho _{SSB} = 3H_{0}^{2}M_{pl}^{2}\Omega _{m} (1+z_{SSB})^{3}$ is the background density at the redshift $z_{SSB}$ of spontaneous symmetry breaking (SSB), $\beta_0$ is a model parameter and $\phi _{0}$ is the symmetry breaking vacuum expectation value (VEV) of the field 
for $\rho _{m} \rightarrow 0$ \footnote{Since $\phi (a) \propto \phi _{0}$, linear perturbations 
do not depend on the VEV value, and we do not need to specify $\phi_0$.}. 
We define $L=H_0/\mu$, and fix $\beta_0=L=1$ to reflect simulated values, leaving only $z_{SSB}$ as a free parameter in our analysis.  

\subsection{Linear Power Spectrum}

\begin{figure*}
\begin{center}

\includegraphics[width=3.2in]{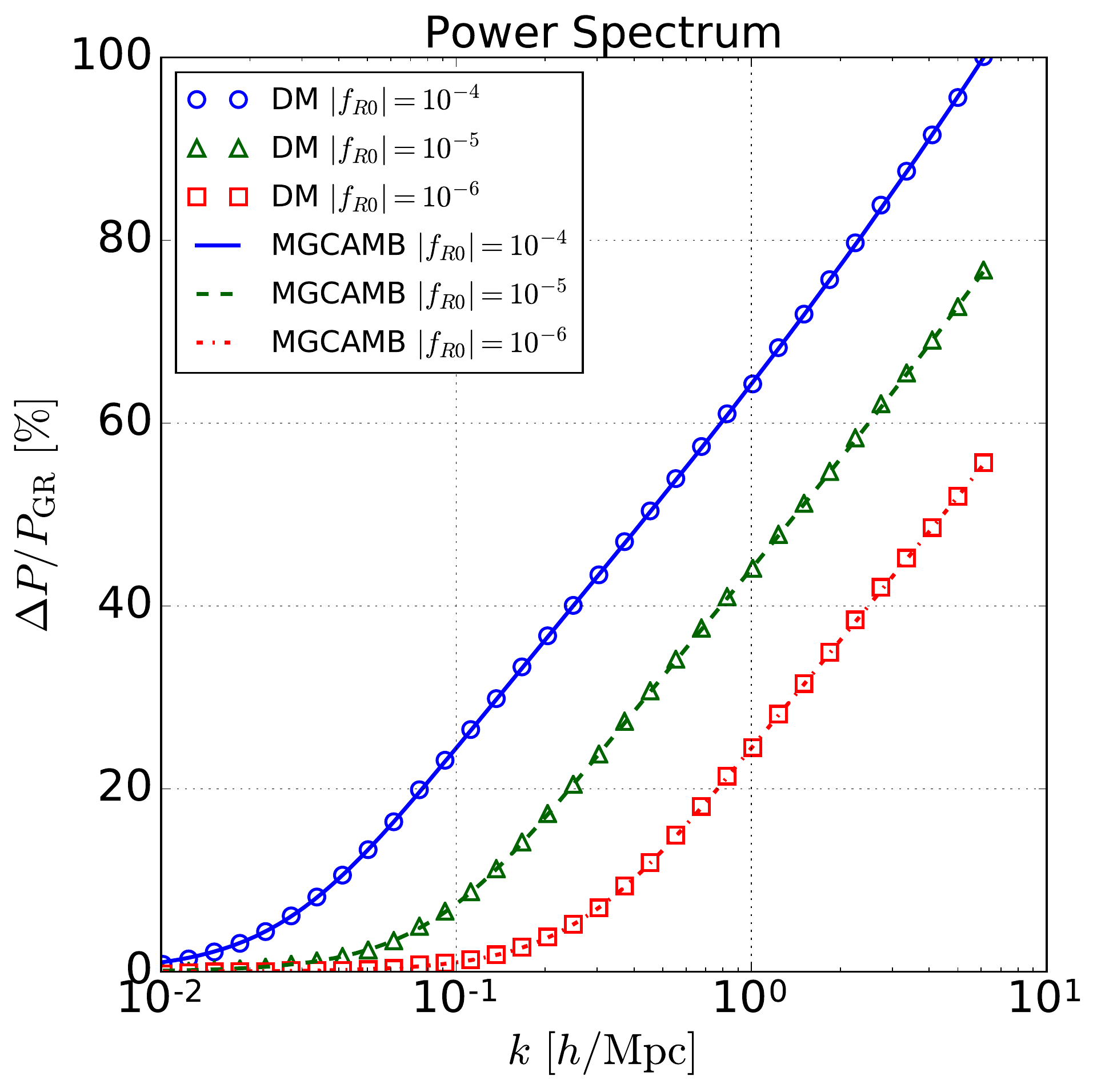}
\includegraphics[width=3.2in]{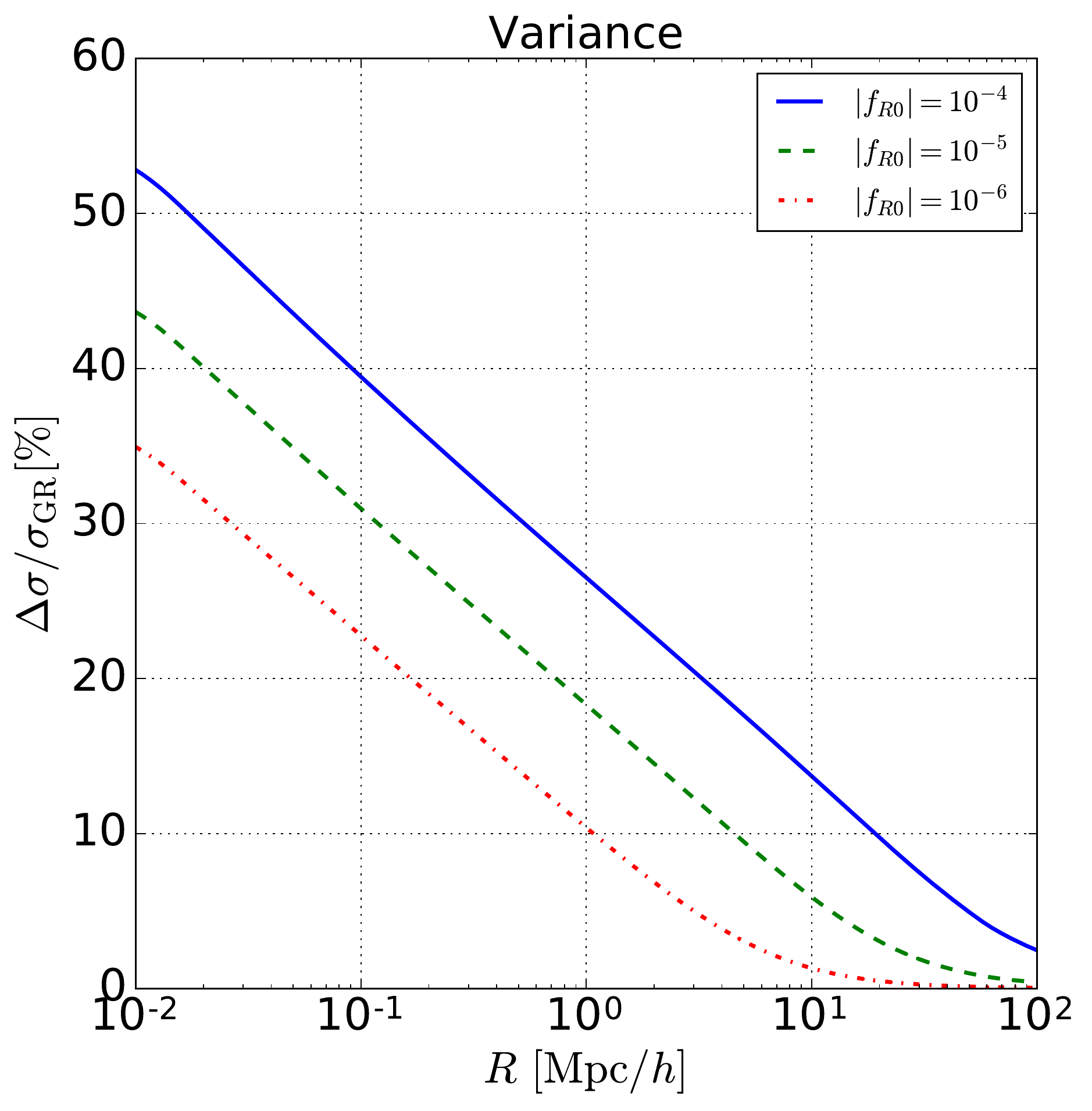}
\caption{ ({\it Left }): Relative percent deviation in the linear matter power spectrum $P(k)$ at $z=0$ of $f(R)$ modified gravity 
with respect to the GR spectrum $P_{GR}(k)$ in $\Lambda$CDM. Results are shown for spectra obtained from \texttt{MGCAMB} 
(lines) as well as from evolving Eq.~\eqref{delta_lin} for dark matter perturbations (open dots), 
for $|f_{R0}| = 10^{-4}$ (blue solid line and circles), $10^{-5}$ (green dashed line and triangles) and $10^{-6}$  (red dot-dashed line and squares).
({\it Right}): Percent deviation with respect to GR of the mean square density $\sigma(R)=S(R)^{1/2}$ 
smoothed at scale $R$, computed from Eq.~\eqref{sigma} at $z=0$ for the $f(R)$ model. 
In this case, the power spectrum was evaluated from  Eq.~\eqref{delta_lin}. 
}
\label{power_spectrum}
\end{center}
\end{figure*}

	We start by defining the 
	linear density contrast field $\delta (R)$
	smoothed on a scale $R$ around  $\textbf{x}=0$ \footnote{The choice  $\textbf{x}=0$ is irrelevant 
	because of translational invariance in a homogeneous Universe, and is 
	used for simplificity here, 
	as we are interested in the behaviour of $\delta$ as a function of 
	scale $R$.}
\begin{equation}
\delta (R) = \int \frac{d^{3}k}{(2\pi)^{3}}\tilde{\delta}(\textbf{k})\tilde{W}(k,R) \,,
\label{window}
\end{equation}
where tildes denote quantities in Fourier space and $W(\textbf{x},R)$ is the 
window function that smooths the original field $\delta(\textbf{x})$ on scale $R$.

	The variance $S(R)=\sigma^2(R)$ of the linear density field can be written as
\begin{equation}
S(R) = \langle |\delta (R)|^2 \rangle = \int \frac{dk}{2\pi ^{2}} k^{2} P(k) |\tilde{W}(k,R)|^{2}\,,
\label{sigma}
\end{equation}	
where $P(k)$ is the linear power spectrum defined via
\begin{equation}
\langle \tilde{\delta}(\textbf{k})\tilde{\delta}(\textbf{k}')\rangle = (2\pi)^{3} \delta _{D}(\textbf{k} - \textbf{k}')P(k)\,,
\end{equation}
and $\delta _{D}(\textbf{k} - \textbf{k}')$ is a Dirac delta function. 
	Clearly the linear power spectrum will play a key role in describing the effects of modified gravity on void properties.  
	For GR computations, we use  \texttt{CAMB} \cite{CAMB} to compute the linear 
	power spectrum.
	For modified gravity, we may use \texttt{MGCAMB} \cite{Hojjati,Zhao}, a modified 
	version of \texttt{CAMB} which  
	generates the linear  spectrum for a number of alternative models, such as the 
	Hu \& Sawicki $f(R)$ model \cite{Hu} in Eq.~\eqref{fr} and others. 
	However it does not compute the linear spectrum 
	for instance for the symmetron model. Therefore we also construct the linear power 
	spectrum independently for an arbitrary gravity theory parametrized by 
	Eqs.~\eqref{delta_lin} and ~\eqref{mu}. 
	
	Our independent estimation of the spectrum is accomplished by evolving Eqs.~\eqref{delta_lin} 
	and \eqref{mu} with 
	parameters from specific gravity theories (e.g. 
	Eq.~\eqref{para_fr} for $f(R)$ and Eq.~\eqref{para_symm} for  
	symmetron models) for a set of initial conditions at matter domination. 
	Since at sufficiently high redshifts 
	viable gravity models reduce to GR, we take initial conditions given by \texttt{CAMB} at 
	high redshifts ($z \approx 100$), when gravity is not yet modified and the Universe is 
	deep into matter domination. We also compute 
	initial conditions for $\dot{\delta}$ numerically by using the $\Lambda$CDM power spectrum at 
	two closeby redshifts, e.g. at $z=99$ and $z=100$. 
	
	The results of using this procedure are shown (open dots) on the left panel of 
	Fig.~\ref{power_spectrum} and compared with the results from \texttt{MGCAMB} (lines) for 
	the Hu \& Sawicki model with $n=1$ and three values of the 
	parameter $|f_{R0}|=10^{-4}, 10^{-5}, 10^{-6}$.        
	We can see that solving 
	Eq.~\eqref{delta_lin} for the power spectrum produces results nearly identical to the full solution from 
	\texttt{MGCAMB} on all scales of interest. The percent level differences may be traced 
	to the fact that the  
	simplified equation solved does not contain information about photons 
	and baryons, but only dark matter. 
	For our purposes, this procedure can 
	be used to compute 
	the linear power spectrum for other modified gravity models that reduce to 
	GR at high redshifts, such as the symmetron model. 

	On the right panel of Fig.~\ref{power_spectrum} we see that the relative difference of $\sigma(R)=S(R)^{1/2}$ for the $f(R)$ model 
	with respect to GR can be significant 
	on the scales of interest ($1$ Mpc/$h < R < 20$ Mpc/$h$). Therefore we expect a 
	similar impact on void properties derived from $\sigma$ and the linear power spectrum.

\subsection{Spherical Collapse}

	Because of the void-in-cloud effect \footnote{The fact that voids inside halos are eventually swallowed and disappear.}, 
	the linearly extrapolated density contrast $\delta _{c}$ for the formation of halos is important in describing the properties of voids
	as both are clearly connected. 
	Within theoretical calculations of the void abundance using the excursion set formalism, $\delta_c$ corresponds to another 
	absorbing barrier, whose equivalent is not present for halo abundance. Therefore calculating $\delta_c$ in the gravity theory 
	of interest gives us 
	important hints into the properties of both halos and voids.
	
	The computation of $\delta_c$ is done similarly to that of the GR case, but using  
	Eqs.~\eqref{delta_eq} and 
	\eqref{delta_lin} with the appropriate modified gravity parameterization $\mu (k,a)$ (GR is recovered with $\mu (k,a)=1$).
	
	Here we followed the procedure described in \cite{Pace}. We start with appropriate initial conditions \footnote{This initial condition is 
	actually determined by a shooting method, evolving the nonlinear Eq.~\eqref{delta_eq} 
	for multiple initial values and checking when collapse happens ($\delta \rightarrow \infty$) 
	at $a=a_{c}$}
	 for $\delta $ and 
	$\dot{\delta} $ and  
	evolve the the linear Eq.~\eqref{delta_lin} until $a_c $.  
	The value of $\delta$ obtained is $\delta _{c}$, the density contrast linearly extrapolated for halo formation 
	at $a=a_c$. In this work, since we only study simulation outputs at $z=0$, we take 
	$a_{c}=1$ in all calculations.
	The only modification introduced by a nontrivial parameterization $\mu (k,a)$ is that the collapse parameters will depend on the scale $k$ of the 
	halo. As mentioned previously, the parameterization of Eq.~\eqref{mu} only takes into account  
	the evolution of the scalar field in the 
	background \footnote{For instance, the scalar field mass in Eq.~\eqref{para_fr} 
	depends only on scale 
	factor $a$, not on the local potential or the environment as would be expected in a full chameleon calculation for $f(R)$.}, and does not account for the
	dependence of the collapse parameters on screening effects. 
	Even though our calculation is approximated, it 
	does approach the correct limits at sufficiently large and small scales.    
	
\begin{figure*}
\begin{center}
\includegraphics[width=3.2in]{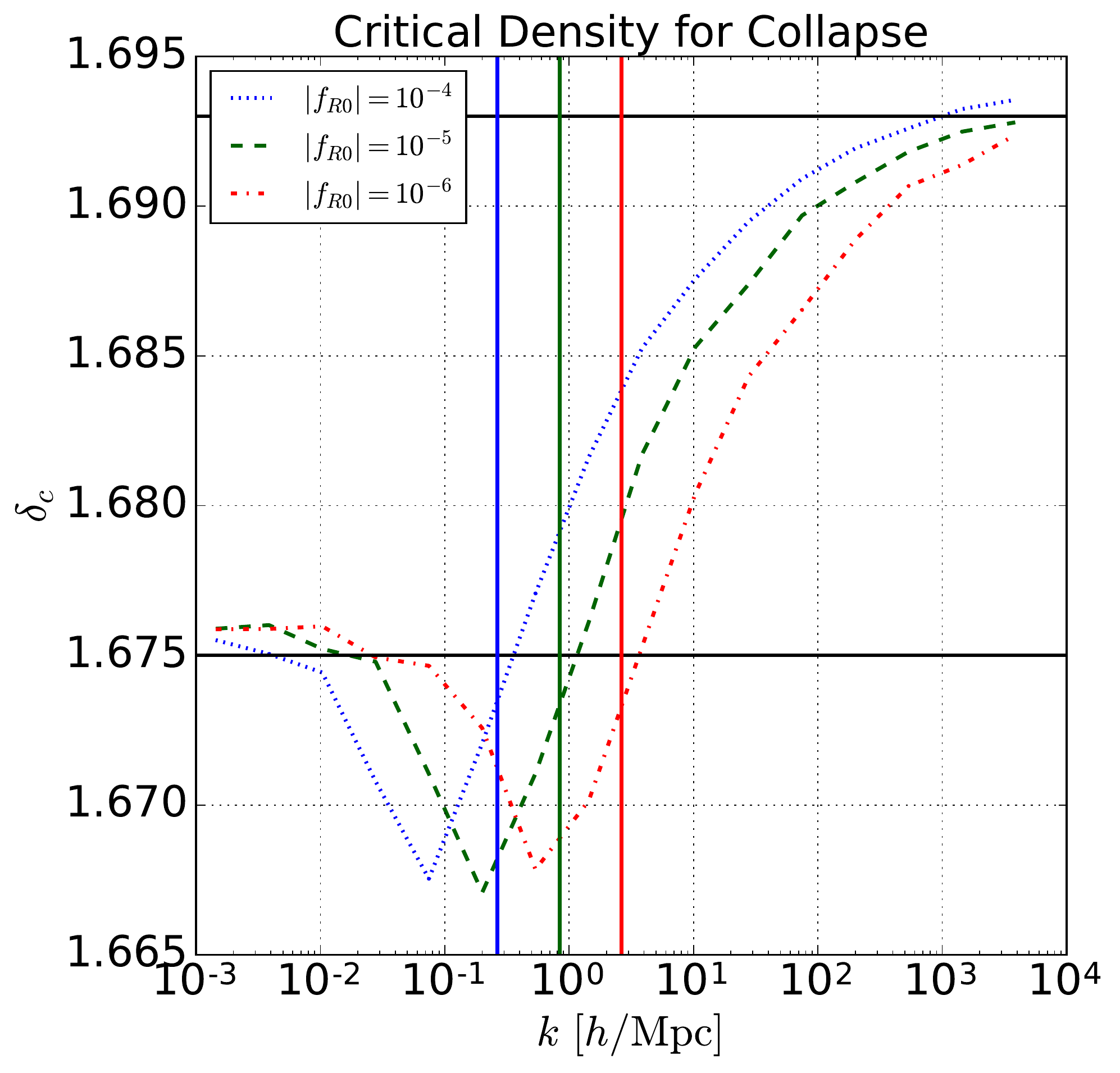}
\includegraphics[width=3.2in]{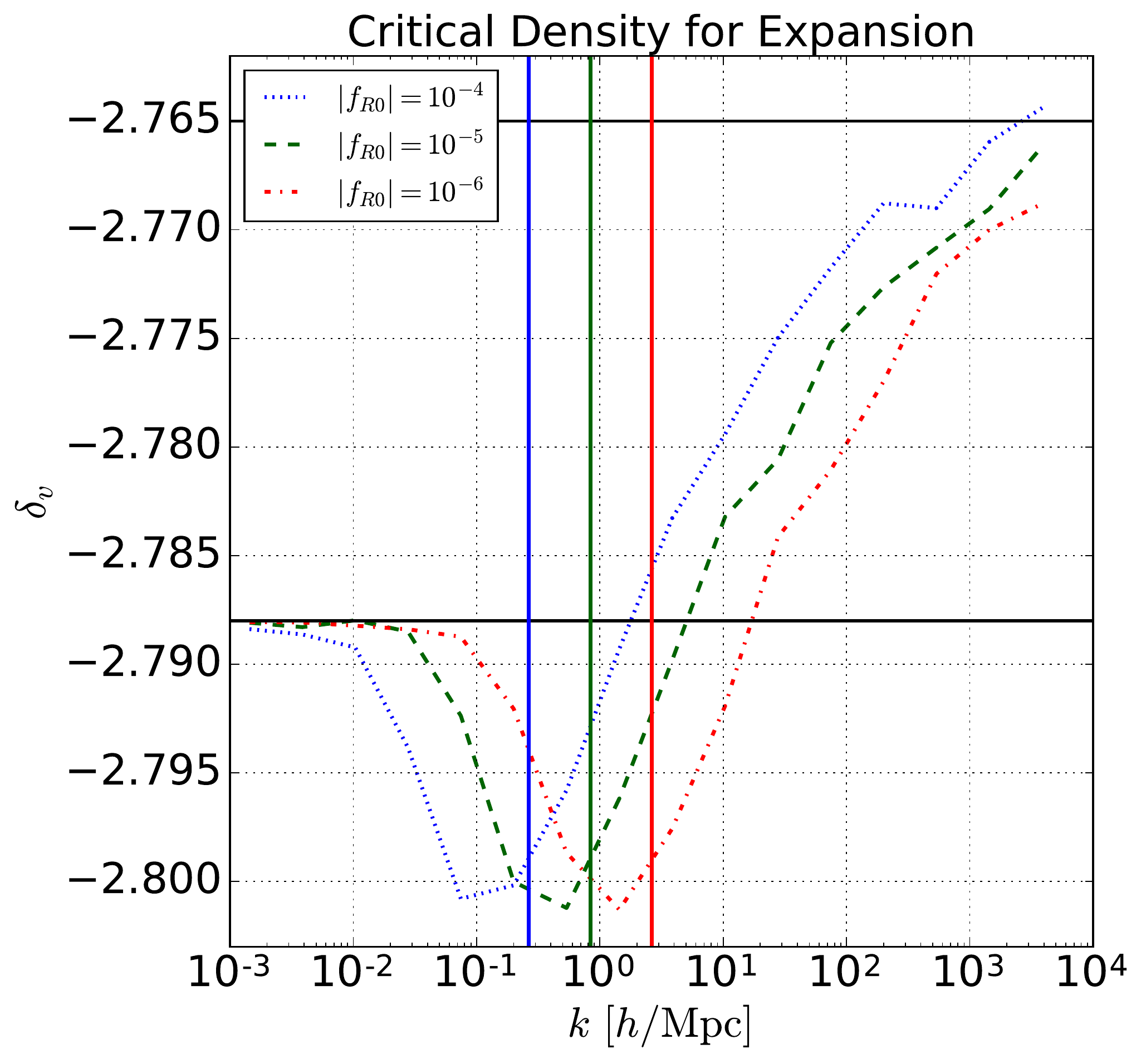}
\caption{({\it Left}): The critical density $\delta_c$ for collapse of a halo at $z=0$ as a function of  
halo scale $k$ in $f(R)$ modified gravity parameterized by Eq.~\eqref{para_fr} with $|f_{R0}|=10^{-4}$ (blue dotted line), $10^{-5}$ (green dashed line) and  $10^{-6}$ (red dot-dashed line). The upper horizontal black line is the value expected for the strong field limit ($\mu = 4/3$) and the lower line 
for the weak field limit, i.e. GR ($\mu=1$). The vertical lines indicate the Compton scales for each gravity with the same corresponding line colors. 
({\it Right}): Same for the critical density $\delta_v$ for void formation at $z=0$. 
}
\label{delta_c}
\end{center}
\end{figure*}	

	For a Universe with only cold dark matter (CDM) under GR, the collapse equations can be solved analytically 
	yielding $\delta_c=1.686$. For a $\Lambda$CDM Universe, still within GR, $\delta_c$ changes to a 
	slightly lower value, whereas for stronger gravity it becomes slightly larger. 
	In Fig.~\ref{delta_c} we show $\delta_c$ as function of scale for the $f(R)$ model. 
	The value of $\delta _{c}$ starts at 
	its $\Lambda$CDM value $\delta_c=1.675$ on scales larger than 
	the Compton scale ($k/a \ll m$; weak field limit where $\mu\approx 1$) and approaches 
	the totally modified value $\delta_c=1.693$ on smaller scales ($k/a \gg m$; strong field 
	limit where $\mu\approx 1+2\beta^2=4/3$) where the modification to the strength of 
	gravitational force is maximal. These values were computed at the background cosmology described in 
	\S~\ref{sec:void_sim}. They are similar to those of \cite{Schmidt09}, though the cosmology is slightly different. 
	Note that $\delta _{c}$ reaches its strong field limit faster for 
	larger values of $|f_{R0}|$ (value of the extra scalar field today), as expected. 
	In the approximation of Eq.~\eqref{delta_eq}, $\delta _{c}$ varies with $k$ less than in the full collapse \cite{Kopp, Borisov}, 
	indicating that the no-screening approximation may not be sufficient. As a full exact calculation is beyond 
	the scope of this work and given that $\delta_c$ does not change appreciably, in our abundance models we will fix 
	$\delta_c$ to its $\Lambda$CDM value and encapsulate modified gravity effects  on the linear 
	power spectrum and on
	other model parameters. 

\subsection{Spherical Expansion}

	We now compute $\delta _{v}$, the analog of $\delta _{c}$ for voids, i.e. the density contrast linearly extrapolated to today 
	for the formation of a void. 
	We follow a procedure similar to spherical collapse, but in this case the initial values for 
	$\delta _{i}$ are negative. We also set a 
	criterium in the nonlinear 
	field $\delta$ for the formation of a void to be \footnote{$\delta _{sc}=-0.8$  is the density contrast in which shell-crossing ($sc$) 
	occurs in an Einstein-de-Sitter (EdS) Universe \cite{Jennings}.}  $\delta _{sc} = -0.8$ 
	or equivalently 
	$\Delta_{sc}=1+\delta_{sc}=0.2$ \cite{Jennings}.
	This quantity is somewhat the analogue for voids of the virial overdensity $\Delta_{vir}\approx 180$ for halo formation in 
	an  Einstein-de-Sitter (EdS) Universe. 
	Despite the value of $\Delta_{vir}$ being only strictly appropriate for an EdS Universe, halos are often defined 
	with this overdensity or other arbitrary values that may be more appropriate for specific observations. 
	Similarly, $\delta_{sc}=-0.8$ is only strictly appropriate for shell-crossing in an EdS Universe. Here we will employ 
	$\delta_{sc}=-0.8$, but we should keep in mind that this is an arbitrary definition of 
	our spherical voids. 
	When we fix this criterium for void formation we also fix the factor by which the 
	void radius $R$ expands with respect to its linear theory radius $R_L$. This factor is given by 
	$R/R_L=(1+\delta _{sc})^{-1/3} = 1.717$ \cite{Jennings}, 
	and comes about from mass conservation throughout the expansion. 
        Differently from halos, voids are not virialized structures and continue to expand faster than the background. 	
	Again environmental dependences are not incorporated in our computations as these values will depend 
	only on scale factor $a$ and the scale $k$ or size of the void.

	The right panel of Fig.~\ref{delta_c} displays the behaviour of $\delta _{v}$ as a function of $k$, which is very similar to that 
	of $\delta _{c}$.  
	This is important when modelling the absorbing barriers used for 
	evaluating the void abundance distribution function. Again the values of $\delta _{v}$ vary with $k$ less than in the full calculation 
	\cite{Clampitt}. 
	
\begin{table}
\centering
\caption{Critical densities for the spherical collapse and expansion in the weak and strong field limits in $f(R)$ gravity. }
\begin{tabular}{c|c|c|c}
\hline 
Limit & $\mu$ & $\delta _{c}$  & $\delta _{v}$ \\ 
\hline                           
Weak Field & 1 & 1.675 & -2.788 \\
Strong Field & 4/3 & 1.693 & -2.765 \\
\hline   
\end{tabular}
\label{tab:deltac}
\end{table}

	In Table~\ref{tab:deltac}, we show the values of $\delta_c$ and $\delta_v$ 
	in the weak and strong field limits of $f(R)$ gravity. We see that the parameters are not very much affected by the strong change 
	in gravity ($1\%$ for $\delta _{c}$ and $0.8\%$ for $\delta _{v}$) compared with the change 
	induced in the linear variance (see Fig.~\ref{power_spectrum}). Even though these collapse/expansion parameters 
	come inside exponentials in the modeling of void abundance, these results indicate that the main 
	contribution from gravity effects appear in the linear spectrum.
	
	The spherical collapse and expansion calculations can be performed similarly for the symmetron model, with the 
	appropriate change in the expression for the mass and coupling of the scalar field, as given by the 
	Eq.~\eqref{para_symm}. For $f(R)$ gravity the change in parameters 
	does not seem to be relevant and we fix these parameters to their $\Lambda$CDM values. In order to treat 
	both gravity models in the same way, we do the same for the symmetron model. Therefore we do not show explicit 
	calculations of $\delta_c$ and $\delta_v$ for symmetron.

\section{Void Abundance Function} \label{sec:void_dist}

	We now compute the void abundance distribution function as a function of void size 
	using an extended Excursion Set formalism \cite{Maggiore1}, which  
	consists in solving the Fokker-Planck equation with appropriate boundary conditions \footnote{This procedure is valid 
	when the barrier (boundary conditions) is linear in $S$ and the random walk motion is Markovian.}.
	
	Differently from the halo description, for voids it is necessary to use two boundary conditions, because of the 
	void-in-cloud effect \cite{Sheth1}. In this case we use two Markovian stochastic barriers with linear dependence in the 
	density variance $S$, which is a simple generalization from the conventional problem with a constant barrier. 
	The barriers can be described statistically as
\begin{eqnarray}
\langle B_{c}(S) \rangle &=& \delta _{c} + \beta _{c} S\,, \nonumber \\
\langle B_{c}(S)B_{c}(S')\rangle &=& D_{c} \min(S,S')\,, \nonumber \\
\langle B_{v}(S)\rangle &=& \delta _{v} + \beta _{v} S\,, \nonumber \\
\langle B_{v}(S)B_{v}(S')\rangle &=& D_{v} \min(S,S')\,,
\label{barries}
\end{eqnarray}
where $B_{c}(S)$ is the barrier associated with halos and $B_{v}(S)$ the barrier associated with voids. Notice that 
the two barriers are uncorrelated, i.e. $\langle B_{c}(S)B_{v}(S') \rangle=0$.
Here $\beta_c$ describes the linear relation between the mean barrier and the variance $S$, 
$\delta_{c,v}$ is the mean barrier as $S\rightarrow 0$ ($R\rightarrow \infty$), and $D_{c,v}$ 
describes the barrier diffusion coefficient. 

	As we consider different scales $R$, the smoothed density field $\delta(R)$ performs a random walk with respect 
	to a \emph{time coordinate} $S$, and we have \footnote{This occurs when the window function in Eq.~\eqref{window} $S$ is sharp in $k$-space. 
	For a window that is sharp in real space the motion of $\delta$ is not Markovian and the second equation in \eqref{delta} is not true. 
	In that case a more sophisticated method is necessary (see \cite{	Maggiore1} for details), and the solution presented here 
	represents the zero-order approximation for the full solution.}
\begin{eqnarray}
\langle \delta (S)\rangle &=& 0\,, \nonumber \\
\langle \delta (S) \delta (S')\rangle &=& \min(S,S')\,.
\label{delta}
\end{eqnarray}

	The field $\delta$ satisfies a Langevin equation with white noise and therefore the probability density $\Pi(\delta,S)$ to find the 
	value $\delta$ at variance $S$ is a solution of the Fokker-Planck equation
\begin{equation}
\frac{\partial \Pi}{\partial S} = \frac{1}{2}\frac{\partial ^{2} \Pi}{\partial \delta ^{2}}\,,
\label{FP1}
\end{equation}
with boundary conditions
\begin{equation}
\Pi(\delta = B_{c}(S),S) = 0 \quad \mbox{and} \quad \Pi(\delta = B_{v}(S),S) = 0\,,
\label{boundary1}
\end{equation}
and  initial condition
\begin{equation}
\Pi(\delta ,S=0) = \delta _{D}(\delta)\,,
\label{initial1}
\end{equation}
where $\delta_{D}$ is a Dirac delta function and notice that $S\rightarrow 0$ 
corresponds to void radius $R\rightarrow \infty$. 
In order to solve this problem, it is convenient to introduce the 
variable \cite{Achitouv2}
\begin{equation}
Y(S) = B_{v}(S) - \delta (S) \,.
\end{equation}

Making the {\it simplifying} assumption that $\beta \equiv \beta _{c} = \beta _{v}$ \footnote{Notice that $\beta$ here should not be 
confused with the coupling between matter and the extra scalar in Eq.~\eqref{mu}} 
and using the fact that all variances can be added in quadrature, the Fokker-Planck Eq.~\eqref{FP1} becomes 
\begin{equation}
\frac{\partial \Pi}{\partial S} =  -\beta \frac{\partial \Pi}{\partial Y} + \frac{1+D}{2}\frac{\partial ^{2} \Pi}{\partial Y ^{2}}
\label{FP2}
\end{equation}
where $D=D_{v}+D_{c}$.

	 We define $\delta _{T} = |\delta _{v}| + \delta _{c}$ and notice that $\delta (S) = B_{v} (S)$ implies $Y(S) = 0$, $\delta (S) = B_{c}(S)$
	 implies  $Y(S) = -\delta _{T}$ (only occurs because we set $\beta _{c} = \beta _{v}$) and $\delta (0) = 0$ implies $Y(0) = \delta _{v} $. 
	 Therefore the boundary conditions become
\begin{equation}
\Pi (Y =0,S) = 0 \quad \mbox{and} \quad \Pi (Y =-\delta _{T},S) = 0\,,
\label{boundary2}
\end{equation}
and the initial conditions 
\begin{equation}
\Pi (Y,0) = \delta _{D}(Y-\delta _{v})\,.
\end{equation}

	Rescaling the variable $Y \rightarrow \tilde{Y} = Y/\sqrt{1+D}$ and factoring the solution in the form $\Pi(\tilde{Y},S) = U(\tilde{Y},S)\exp [c(\tilde{Y} - cS/2 - \tilde{Y}_{0})]$ where $c=\beta /\sqrt{1+D}$ and $\tilde{Y}_{0} = \delta _{v}/\sqrt{1 + D}$. The function $U(\tilde{Y},S)$ obeys a Fokker-Planck equation like Eq.~\eqref{FP1}, for which the solution is known \cite{Sheth1}. Putting it all together the probability distribution function becomes
\begin{eqnarray}
\Pi (Y,S) &=& \exp \left[  \frac{\beta}{1+D}\left(Y - \frac{\beta S}{2} - \delta _{v}\right) \right] \nonumber \\
&&\hspace*{-55pt} \times \sum _{n=1} ^{\infty} \frac{2}{\delta _{T}} \sin \left( \frac{n \pi \delta _{v}}{\delta _{T}} \right) \sin \left( \frac{n\pi}{\delta _{T}} Y \right) \exp \left[ - \frac{n^{2} \pi ^{2} (1+D)}{2 \delta _{T} ^{2}} S \right]\,. \nonumber \\
\end{eqnarray}

	The ratio of walkers that cross the barrier $B_{v}(S)$ is then given by

\begin{equation}
\mathcal{F}(S) = \frac{\partial}{\partial S} \int _{\infty} ^{0} dY \Pi (Y,S) =  \frac{1+D}{2} \left. \frac{\partial \Pi}{\partial Y} \right|_{Y=0}\,,
\end{equation}
where we used the modified Fokker-Planck equation Eq.~\eqref{FP2} and the first boundary 
condition from Eq.~\eqref{boundary2}.
The void abundance function, defined as $f(S) = 2S\mathcal{F}(S)$, for this model is then given by
\begin{eqnarray}
f(S) &= &2 (1+D) \exp \left[ -\frac{\beta ^{2} S}{2(1+D)} + \frac{\beta \delta _{v}}{(1+D)} \right]\nonumber \\
&\times& \sum _{n=1} ^{\infty} \frac{n\pi }{\delta _{T}^{2}} S \sin \left( \frac{n \pi \delta _{v}}{\delta _{T}} \right) \exp \left[- \frac{n^{2} \pi ^{2} (1+D)}{2 \delta _{T} ^{2}} S \right] \nonumber \\
\label{my}
\end{eqnarray}

	There are four important limiting cases to consider: 
	
\begin{itemize}

\item $D=\beta =0$: This is the simplest case of two static barries. The expression in this case was first obtained in \cite{Sheth1} and compared to simulations in \cite{Jennings}. It is given by
\begin{eqnarray}
f_{D=\beta =0}(S) &=& 2 \sum _{n=1} ^{\infty} \frac{n\pi }{\delta _{T}^{2}} S \sin \left( \frac{n \pi \delta _{v}}{\delta _{T}} \right)\nonumber \\
&\times & \exp \left(- \frac{n^{2} \pi ^{2} }{2 \delta _{T} ^{2}} S \right)\,,
\label{2SB}
\end{eqnarray}
This is one of the functional forms tested in this work and the only case with no free parameters. We refer to this case as that of 
2 static barriers (2SB). 

\item $D=0$ and $\beta \ne 0$: This case considers that the barriers depend linearly on $S$ but are not difusive. In this case the expression is given 
by
\begin{eqnarray}
f_{D=0}(S) &=& 2 e^{-\frac{\beta ^{2} S}{2}} e^{\beta \delta _{v}} \sum _{n=1} ^{\infty} \frac{n\pi }{\delta _{T}^{2}} S \sin \left( \frac{n \pi \delta _{v}}{\delta _{T}} \right)\nonumber \\
&\times & \exp \left(- \frac{n^{2} \pi ^{2} }{2 \delta _{T} ^{2}} S \right)
\end{eqnarray} 
	This expression recovers Eq.~(C10) from \cite{Sheth1}. Note that these authors define the barrier with a negative slope, 
	therefore our $\beta$ is equal to their $-\beta$, but $\delta _{v} <0$ in our case;  

\item $\beta = 0$ and $D \ne 0$: Here we have a barrier that does not depend on $S$ but which is diffusive. In this case we have
\begin{eqnarray}
f_{\beta = 0}(S) &=& 2 (1+D) \sum _{n=1} ^{\infty} \frac{n\pi }{\delta _{T}^{2}} S \sin \left( \frac{n \pi \delta _{v}}{\delta _{T}} \right) \nonumber \\ &\times & \exp \left(- \frac{n^{2} \pi ^{2} (1+D)}{2 \delta _{T} ^{2}} S \right)
\end{eqnarray} 
	This expression is the same as the original formula from \cite{Sheth1}, but changing $S \rightarrow (1+D)S$ or 
	$(\delta _{v}, \delta _{v}) \rightarrow (\delta _{v}, \delta _{c})/\sqrt{1+D}$, as expected when the constant barrier becomes diffusive \cite{Maggiore2};\\

\item {\it Large void radius}: As discussed in \cite{Sheth1} and \cite{Jennings}, for large radii $R$ the void-in-cloud effect is not 
important  as we do not expected to find big voids inside halos. In others words, when $S \rightarrow 0  ( R\rightarrow \infty )$ the abundance becomes 
equal to that of a one-barrier problem. Even though we do not attempt to properly consider the limit 
of Eq.~\eqref{my} when 
$S\rightarrow 0$, this expression can be directly compared to the function of the problem with one 
linear diffusive barrier (1LDB), given by \cite{Corasaniti}
\begin{equation}
f_{1{\rm LDB}}(S) = \frac{|\delta _{v}|}{\sqrt{S(1+D_v)}}\sqrt{\frac{2}{\pi}} \exp \left[- \frac{(|\delta _{v}| + \beta_v S)^{2}}{2S(1+D_v)} \right]
\label{One_barier}
\end{equation}

\end{itemize}

\begin{figure}
\begin{center}
\includegraphics[width=3.2in]{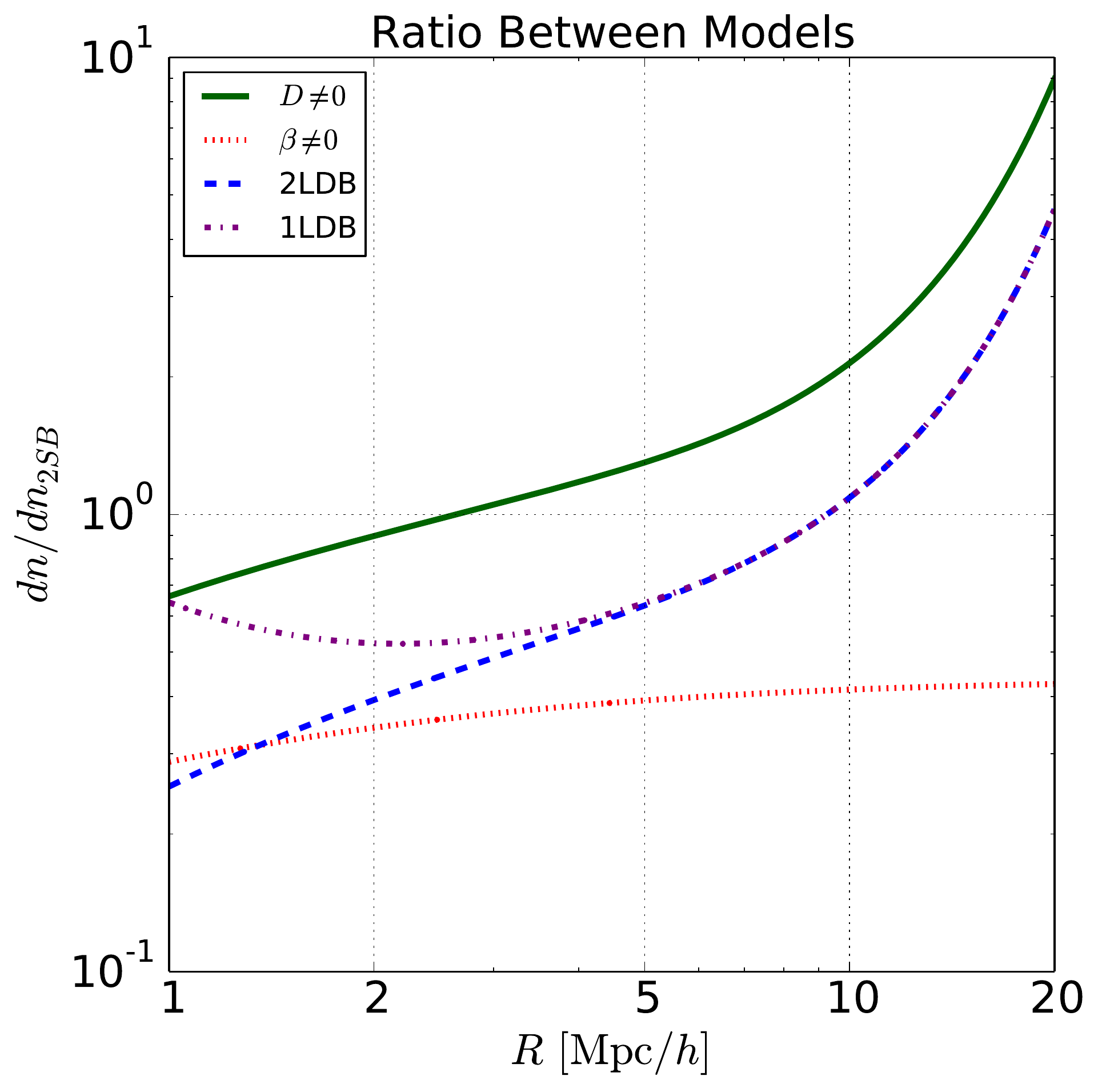}
\caption{Ratio of multiple models for void abundance relative to the model with two static barriers (2SB) Eq.~\eqref{2SB} ($\beta = D = 0$).
We show models with only $D\neq0$ (green solid line), with only $\beta \neq 0$ (red dotted line), 
the 1LDB model (purple dotted-dashed line) and the 2LDB model (blue dashed line). 
The latter two cases are the main models considered in this work and differ only at small 
radii ($R\lesssim 4$ Mpc$/h$), as a manifestation of the void-in-cloud effect. }
\label{comp}
\end{center}
\end{figure}	

	In Fig.~\ref{comp}, we compare the void abundance from multiple cases by taking their ratio 
	with respect to the abundance of the 2SB model. The abundance of the model with 
	$D\neq 0$ is
	substantially higher than 2SB, whereas that of the model with $\beta \neq 0$ is 
	significantly lower.  
	The cases with two linear diffusive barriers (2LDB) Eq.~\eqref{my} 
	and one linear diffusive barrier (1LDB) Eq.~\eqref{One_barier} are the main models 
	considered in this work. The void abundance of the 1LDB and 2LDB models  
	are nearly identical for $R>4$ Mpc/$h$, when the same values of $\beta$ and $D$ are used.
        Table~\ref{tab:models} summarizes the properties of the three main models considered and how they generalize each other. 
	
	\begin{table}
\centering
\caption{Abundance models for  voids considered in this work. Voids require two barriers to avoid the void-in-cloud effect.}
\begin{tabular}{l|c|c|c}
\hline 
Model & Barriers & Nonzero Params & Equation \\ 
\hline                           
2SB &  2 (static)  &$\delta_c$, $\delta_v$                          & Eq.\eqref{2SB} \\
1LDB & 1 (linear+diffusive) &$\delta_v$, $\beta_v$, $D_v$                       & Eq.\eqref{One_barier} \\
2LDB \footnote{ For 2LDB, $\beta=\beta_c=\beta_v$ and $D=D_c+D_v$.} & 2 (linear+diffusive)  &$\delta_c$, $\delta_v$, $\beta$, $D$     & Eq.\eqref{my} \\
\hline   
\end{tabular}
\label{tab:models}
\end{table}

	Given the ratio of walkers that cross the barrier $B_{v}(S)$ with a radius given by $S(R)$, the number density 
	of voids with radius between $R_L$ and $R_L+dR_L$ in linear theory is given by
\begin{equation}
\frac{dn_{L}}{d\ln R_{L}} = \left. \frac{f(\sigma)}{V(R_{L})} \frac{d \ln \sigma ^{-1}}{d \ln R_{L}} \right|_{R_{L}(R)}
\end{equation}
where the subscript $L$ denotes linear theory quantities, $V(R_L)$ is the volume of the spherical void of linear radius $R_{L}$ and recall $S=\sigma ^{2}$.

	Whereas for halos the number density in linear theory is equal to the final nonlinear number density, for voids this is not the case. 
	In fact, Jennings et al. \cite{Jennings} shows that such criterium produces nonphysical void abundances, in which the volume fraction of the Universe occupied by 
	voids becomes larger than unity. Instead, to ensure that the void volume fraction 
	is physical (less than unity) the authors of \cite{Jennings} impose that the volume density is the conserved quantity when going from the 
	linear-theory calculation to the nonlinear abundance. Therefore, when a void expands from $R_{L} \rightarrow R$ it combines
	with its neighbours to conserve volume and not number. This assumption is quantified  by the equation  
\begin{equation}
V(R)dn = \left. V(R_{L})dn_{L} \right|_{R_{L}(R)} \,,
\end{equation}
which implies
\begin{equation}
\frac{dn}{d\ln R} = \left. \frac{f(\sigma)}{V(R)} \frac{d\ln \sigma ^{-1}}{d\ln R_{L}}\frac{d \ln R_{L}}{d\ln R}\right|_{R_{L}(R)} \,,
\label{void_abundance}
\end{equation}
where recall in our case $R=(1+\delta _{sc})^{-1/3}R_L=1.717R_{L}$ is the expansion factor for voids. Therefore 
we have trivially $d\ln R_L/d\ln R =1$ above. 

	The expression in Eq.~\eqref{void_abundance} -- referred as the Vdn model -- along with the function in Eq.~\eqref{my} provide the theoretical prediction for the void 
	abundance distribution in terms of void radius, which will be compared to the abundance of spherical voids found in N-body simulations 
	of GR and modified gravity.

\section{Voids from Simulations}\label{sec:void_sim}

\begin{figure*}
\begin{center}
\includegraphics[width=3.2in]{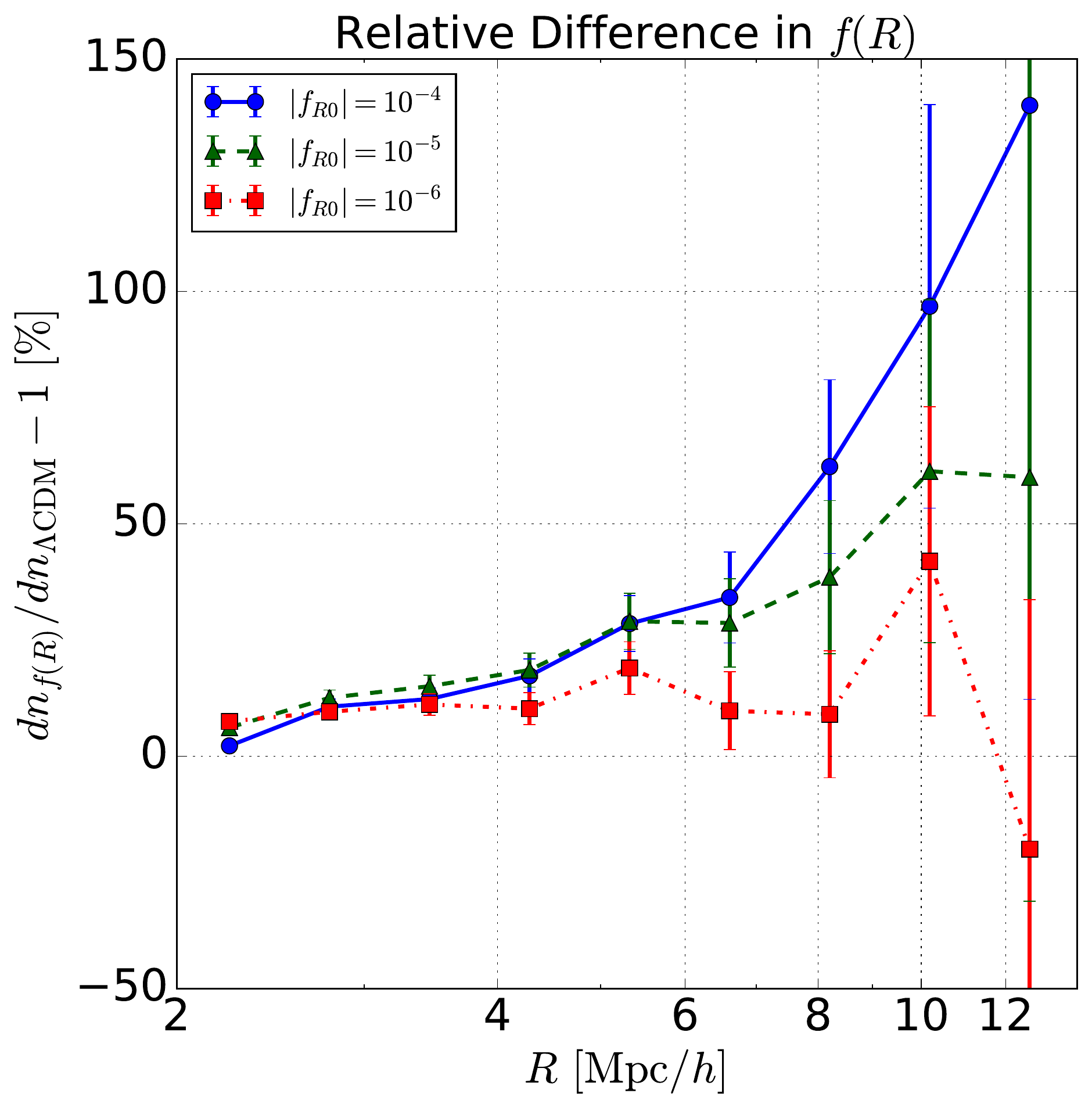}
\includegraphics[width=3.2in]{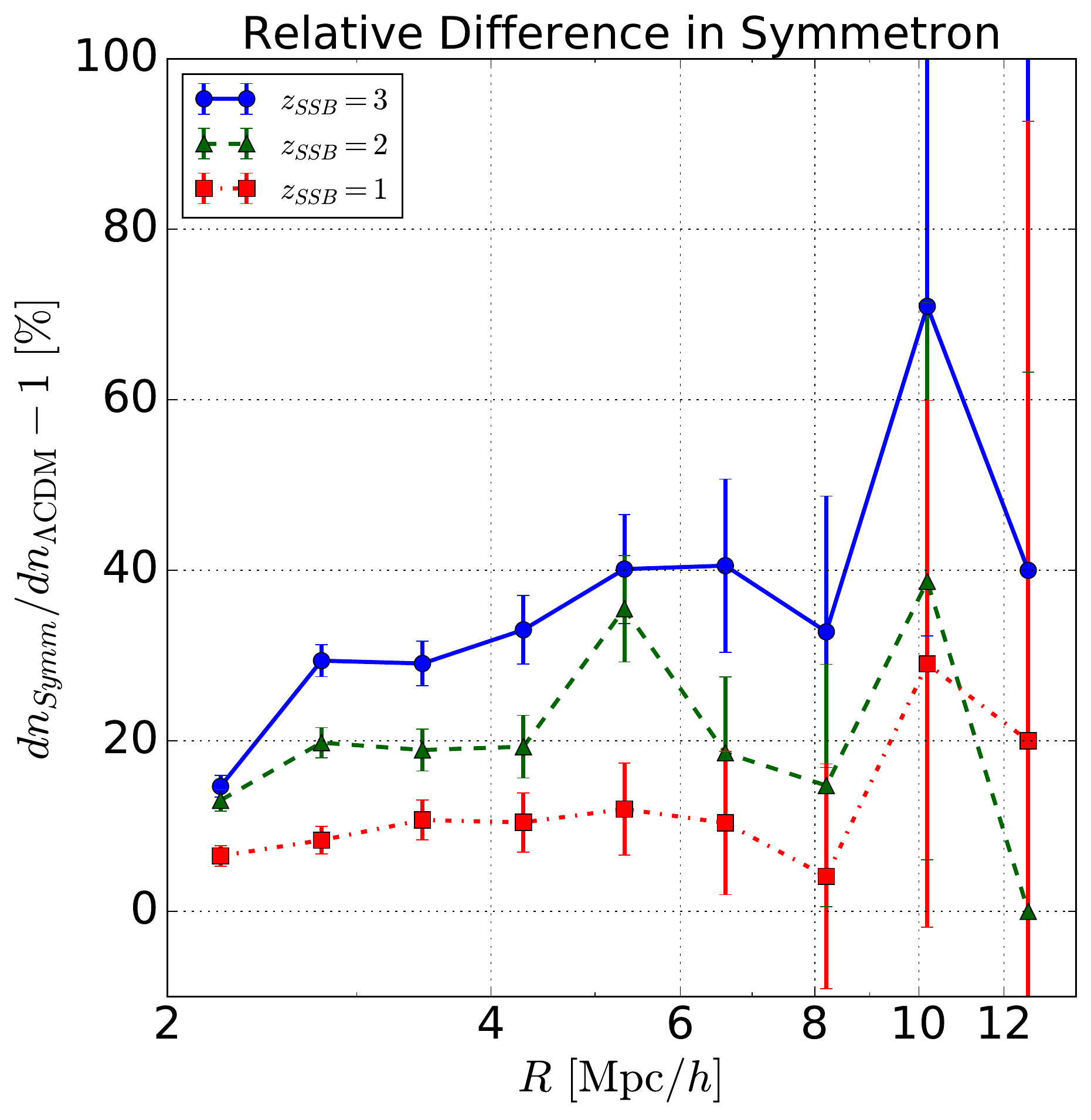}
\caption{Relative difference between void abundance in modified gravity models and in standard GR ($\Lambda$CDM model).
({\it Left}): Relative difference of $f(R)$ theories, for parameters $|f_{R0}| = 10^{-6}$ (red squares with dotted-dashed line), $10^{-5}$ (green triangles with dashed line) and $10^{-4}$ (blue circles with solid line).
({\it Right}): Relative difference of symmetron theories, for parameters $z_{SSB} = 1$ (red squares with dotted-dashed line), $2$ (green triangles with dashed line) and $3$ (blue circles with solid line).}
\label{Diff_Distribution}
\end{center}
\end{figure*}

	We used the N-body simulations that were run with the Isis code \cite{Llinares} for $\Lambda$CDM, $f(R)$ 
	Hu-Sawicki and symmetron cosmological models. For the $f(R)$ case we fixed $n=1$ and considered $|f_{R0}|=10^{-4}$, 
	$10^{-5}$ and $10^{-6}$. For symmetron, we fix $\beta _{0} = 1$ and $L=1$ and used simulations SymmA, SymmB, SymmD, 
	which have $z_{SSB} = 1, 2, 3$ respectively. Each simulation has $512^{3}$ particles in a box of size 
	$256$ Mpc/$h$, and cosmological parameters $(\Omega _{b}, \Omega _{dm}, \Omega _{\Lambda}, \Omega _{\nu}, h, T_{CMB}, n_{s}, \sigma _{8}) = (0.045, 0.222, 0.733, 0.0, 0.72, 2.726 {\rm K}, 1.0, 0.8)$. 
	These represent the baryon density relative to critical, dark matter density, effective cosmological constant density, neutrino density, 
	Hubble constant, CMB temperature, scalar spectrum index and spectrum normalization. 
	The normalization is actually fixed at high redshifts, so that 
	$\sigma _{8}=0.8$ is derived for the $\Lambda$CDM simulation, but is larger for 
	the modified gravity simulations. In terms of spatial resolution, seven levels of refinement were employed on top of a uniform grid with 512 nodes per dimension.  This gives an effective resolution of of 32,678 nodes per dimension, which corresponds to 7.8 kpc/$h$.  The particle mass is $9.26\times 10^9 M_{\odot}/h$.

We ran the {\tt ZOBOV} void-finder algorithm \cite{Neyrinck} -- based on Voronoi tessellation -- 
	 on the simulation outputs 
	 at $z=0$ in order to find underdense regions and define voids, and compared our findings 
	 to the Vdn model of Eq.~\eqref{void_abundance} \cite{Jennings} with the various multiplicity 
	 functions $f(\sigma)$ proposed above (2SB, 1LDB and 2LDB models). 

	 First, we used {\tt ZOBOV} to determine the position of the density minima locations within the
	  simulations
	  and rank them 
	  by signal-to-noise S/N significance. Next, we started from the minimum density point of 
	  highest significance and grew a sphere around this point, adding one
	   particle at a time in each step, until the overdensity $\Delta=1+\delta$ enclosed within the sphere 
	   was $0.2$ times the mean background density of the simulation at $z=0$. Therefore we 
	   defined {\it spherical} voids, which are more closely related to our  
	   theoretical predictions based on spherical expansion.
	  
	We also considered growing voids around the {\it center-of-volume} from the central Voronoi 
	zones. The center-of-volume is defined similarly to the center-of-mass, but each particle position 
	is weighted by the volume of the Voronoi cell enclosing the particle, 
	instead of the particle mass. Using the center-of-volume produces results very similar to the 
	previous prescription, so we only present results for the centers fixed at the density minima. 
	
	In Fig.~\ref{Diff_Distribution} we compare the void abundance inferred from simulations for the 
	three $f(R)$ and the three symmetron 
	theories relative to the $\Lambda$CDM model. Since the differential abundance as a function 
	of void radius is denoted by $dn/d\ln R$, we 
	denote the relative difference between the $f(R)$ and $\Lambda$CDM abundances by 
	$dn_{f(R)}/dn_{\Lambda{\rm CDM}} - 1$ and show 
	the results in terms of percent differences.  The error bars shown here reflect shot-noise 
	from voids counts in the simulation runs. 
	In the $f(R)$ simulation this relative difference is around $100\%$ at radii $R>10$ Mpc$/h$ 
	(for the $|f_{R0}|=10^{-4}$ case). In the symmetron simulation, the difference is around 
	$40\%$ (for the $z_{SSB} = 3$ case), for radii $R\sim8$ Mpc$/h$. This indicates that void 
	abundance is a potentially powerful tool for constraining modified gravity 
	parameters.

\section{Results}\label{sec:res}

\subsection{Fitting $\beta$ and $D$ from Simulations}

	In order to use the theoretical expression in Eq.~\eqref{my} to predict the void abundance 
	we need values for the parameters $\beta$ and $D$. The usual interpretation of $\beta$ is 
	that it encodes, at the linear level, the fact that the true barrier in real cases is not constant. 
	In other words, the contrast density for the void (or halo) formation depends on its size/scale. 	
	This can occur because halos/voids are not perfectly spherical and/or because the expansion 
	(or collapse) intrinsically depends on scale (Birkhoff's theorem is generally not valid in 
	modified gravity). 
	The scale dependency induced by modified gravity can be calculated using our model 
	for spherical collapse (expansion), described in sections II.C and II.D, by fitting a linear relationship between $\delta_{c}$ ($\delta _{v}$) or average barrier 
	$\langle B_c\rangle$ ($\langle B_v \rangle$) as a function of the variance 
	$S(R)$. Here we use $k = 2\pi /R$ to convert wave number to scale $R$.

\begin{figure*}
\begin{center}
\includegraphics[width=3.2in]{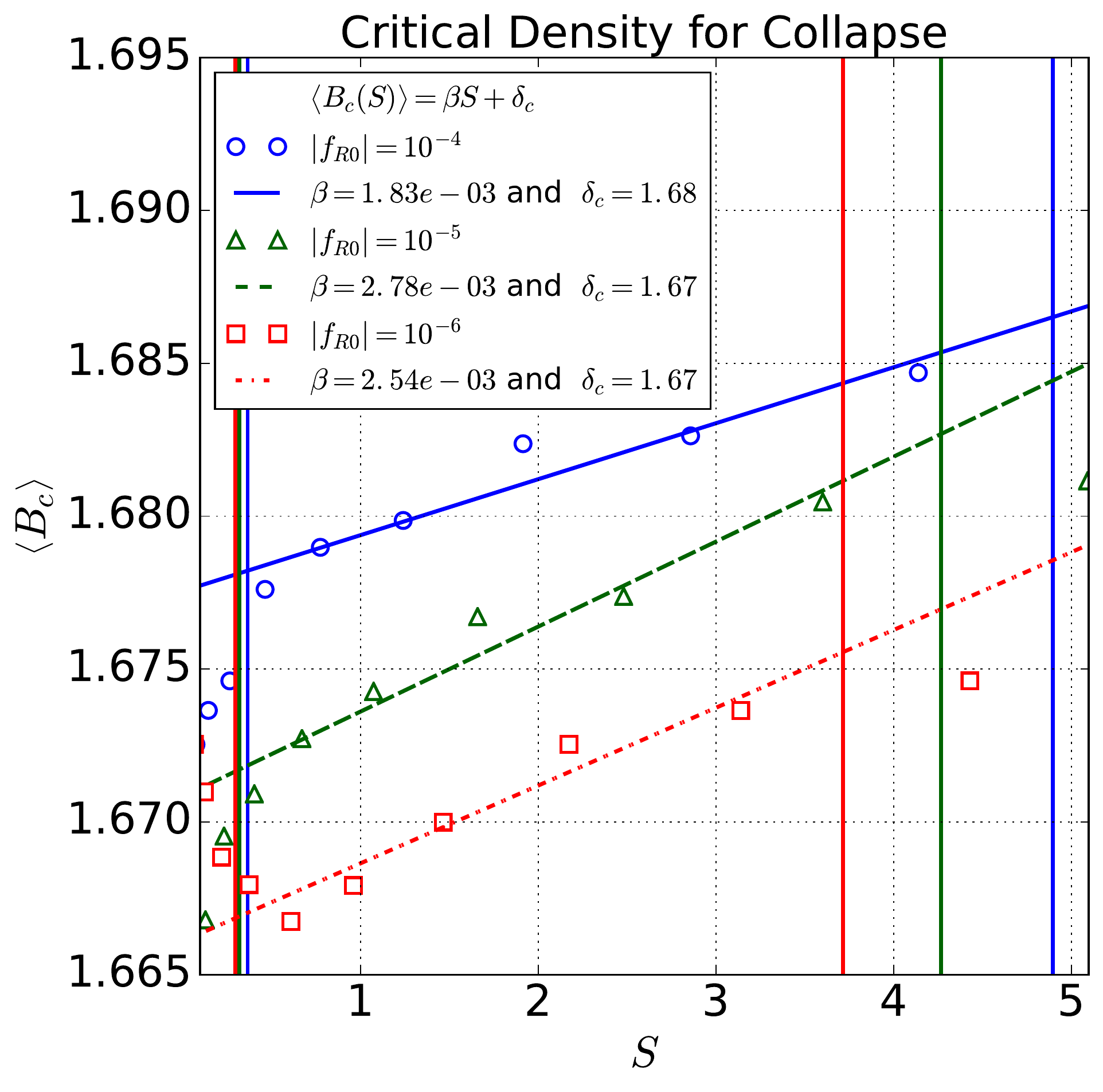}
\includegraphics[width=3.2in]{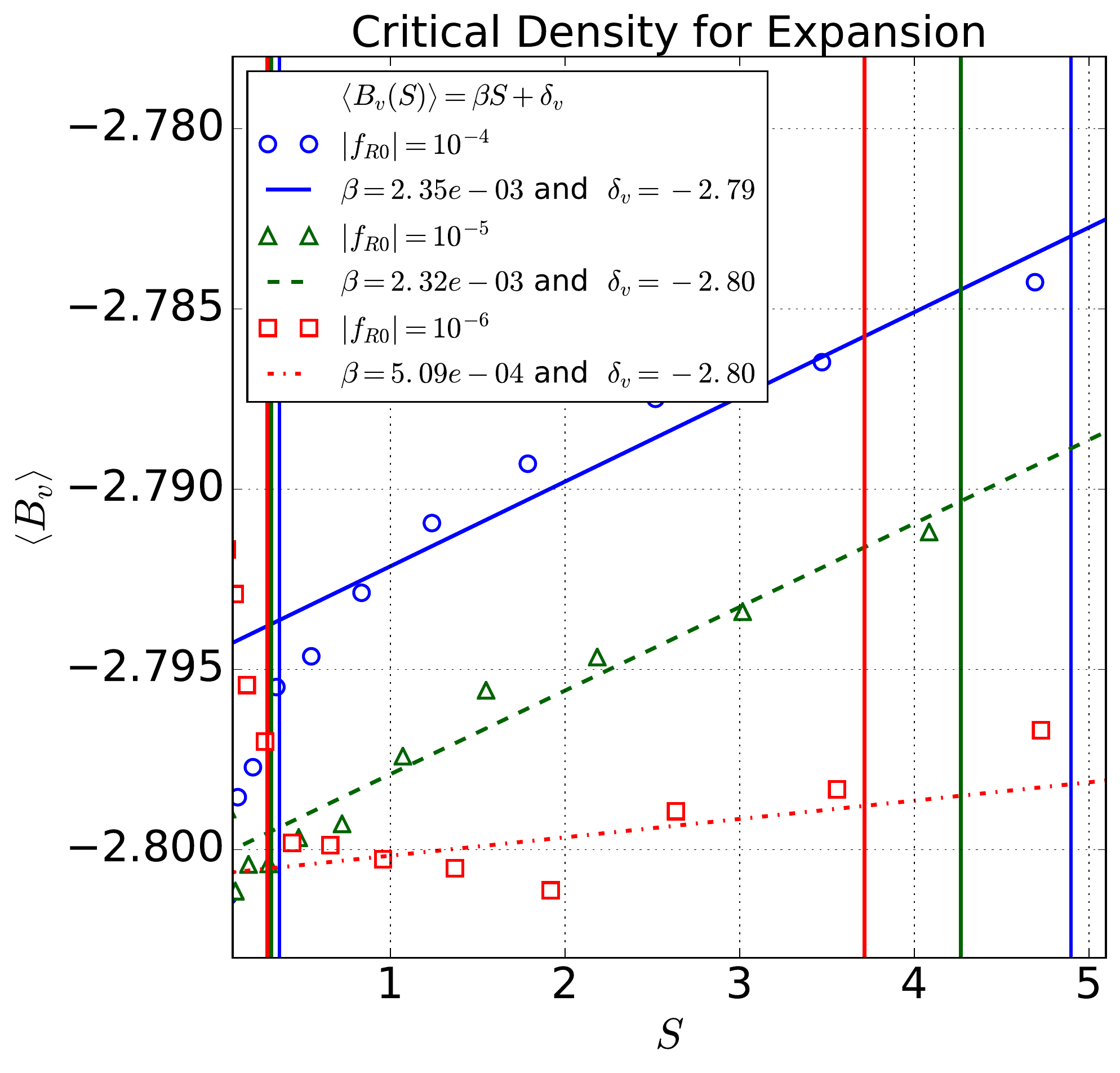}
\caption{ ({\it Left}): Average barrier $\langle B_c \rangle$ for halos 
as a function of variance $S$, 
for the $f(R)$ parameters: $|f_{R0}| = 10 ^{-6}$ (red squares), $10^{-5}$ (green triangles) and $10^{-4}$ (blue circles),
 and corresponding fits for each case in same colors and with dotted-dashed, dashed and solid lines respectively. 
Vertical lines indicate the limits used for the fits, which also correspond to the range of interest for the study of voids in our 
case ($2.0-14.0$ Mpc/$h$).
({\it Right}): Same for the void barrier $\langle B_v \rangle$.}
\label{dc_adjust}
\end{center}
\end{figure*}

        In Fig.~\ref{dc_adjust} we show the average barriers $\langle B_c \rangle$ , $\langle B_v \rangle$
        as functions of variance $S$ for multiple gravity theories, and empirical fits for the parameters 
        $\delta_c, \delta_v, \beta_c, \beta_v$ from Eqs.~\eqref{barries}.
	These fits 
	indicate that the barriers depend weakly on scale in the range of interest. The values of 
	$\delta_c, \delta_v$ are nearly constant and those of 
	$\beta_c, \beta_v$ are of order $10^{-3}$ while the corresponding values for halos in 
	$\Lambda$CDM are of order $10^{-1}$ \cite{Achitouv1}.   
	Even though voids are quite spherical, the small values of $\beta$ indicate that  
	the main contribution to 
	$\beta$ may come from more general aspects of nonspherical evolution. The 
	small fitted values of $\beta$ can also be due to errors induced by the 
	approximations in the nonlinear equation 
	Eq.~\eqref{delta_eq}, which does not capture screening effects of modified gravity. 
	
	Given these issues, and as it is beyond the scope of this work to 
	consider more general collapse models or study the exact modified gravity equations, 
	we will instead keep the values of $\delta_c$ and 
	$\delta_v$ fixed to their $\Lambda$CDM values and treat $\beta$ as a free parameter 
	to be fitted from the abundance of voids detected in the  
	simulations.
	 
	Likewise, the usual interpretation of $D$ is that it encodes stochastic effects of possible 
	problems in our void (halo) finder \cite{Maggiore2}, such as an intrinsic incompleteness or impurity 
	of the void sample, or other peculiarities of the finder, which may even differ from one algorithm to another. 
	Therefore $D$ is also taken as a free parameter in our abundance models.
	
	 We jointly fit for the parameters $\beta$ and $D$ using the voids detected in the N-body 
	 simulations described in \S \ref{sec:void_sim}, with the values of $\delta _{c}$ and $\delta_{v}$ fixed 
	 to their $\Lambda$CDM values (the non-constant barrier introduced by  
	 modified gravity is therefore encoded by $\beta$). 

	 We use the \texttt{emcee} algorithm \cite{emcee} 
	 to produce a Monte Carlo Markov Chain (MCMC) and map the posterior 
	 distribution of these parameters. The results for these fits 
	 using the 2LDB model  Eq.~\eqref{my} the 1LDB model Eq.~\eqref{One_barier} 
	 are shown in Table~\ref{Parameters_all}, for $f(R)$ and symmetron gravity. The 
	 table shows the mean values and 1$\sigma$ errors around the mean,
	  as inferred from the marginalized 
	 posteriors.

\begin{table}
\centering
\caption{Mean values and $1\sigma$ errors for $\beta$ and $D$, fitted 
from void abundance in N-body simulations for GR, $f(R)$ and symmetron 
gravity and for the 1LDB and 2LDB models of void abundance. 
For 1LDB, $\beta=\beta_v$ and $D=D_v$. For 2LDB, $\beta=\beta_c=\beta_v$
and $D=D_c+D_v$.  }
\begin{tabular}{c|c|c|c|c}
\hline 
Gravity &            Parameter    &  Model &                  $\beta $            &                $D$                 \\ 
\hline   &&&& \\                        
GR       &                  -             &    1LDB       & $0.016 ^{0.004} _{0.004}$ & $0.185 ^{0.021} _{0.021}$ \\ &&&& \\  
$f(R)$  & $|f_{R0}|=10^{-6}$  &    1LDB       & $0.029 ^{0.033} _{0.032}$ & $0.168 ^{0.020} _{0.021}$ \\ &&&& \\
$f(R)$  & $|f_{R0}|=10^{-5}$  &     1LDB       & $0.034 ^{0.003} _{0.003}$ & $0.146 ^{0.021} _{0.021}$ \\ &&&& \\
$f(R)$  & $|f_{R0}|=10^{-4}$  &      1LDB     & $0.044 ^{0.003} _{0.003}$ & $0.076 ^{0.021} _{0.021}$  \\ &&&& \\
symmetron & $z_{SSB} = 1$ & 1LDB & $0.010 ^{0.003} _{0.003}$ & $0.150 ^{0.020} _{0.020}$ \\ &&&& \\
symmetron &$z_{SSB} = 2$ & 1LDB & $0.025 ^{0.002} _{0.002}$ & $-0.011 ^{0.016} _{0.017}$ \\ &&&& \\
symmetron &$z_{SSB} = 3$ & 1LDB &$0.034 ^{0.002} _{0.002}$ & $-0.149 ^{0.014} _{0.014}$ \\ 
\hline &&&& \\
GR       &                  -             &    2LDB       & $-0.034 ^{0.002} _{0.002}$ & $0.057 ^{0.014} _{0.014}$ \\ &&&& \\
$f(R)$  & $|f_{R0}|=10^{-6}$  &    2LDB       & $-0.032 ^{0.002} _{0.002}$ & $-0.003 ^{0.012} _{0.011}$ \\ &&&& \\
$f(R)$  & $|f_{R0}|=10^{-5}$  &     2LDB      & $-0.030 ^{0.002} _{0.002}$ & $-0.065 ^{0.011} _{0.012}$ \\ &&&& \\
$f(R)$  & $|f_{R0}|=10^{-4}$  &     2LDB      & $-0.026 ^{0.002} _{0.002}$ & $-0.155 ^{0.010} _{0.010}$\\ &&&& \\ 
symmetron & $z_{SSB} = 1$ & 2LDB & $-0.045 ^{0.002} _{0.002}$ & $0.001 ^{0.012} _{0.012}$\\ &&&& \\
symmetron &$z_{SSB} = 2$ & 2LDB & $-0.032 ^{0.002} _{0.002}$ & $-0.185 ^{0.009} _{0.009}$\\ &&&& \\
symmetron &$z_{SSB} = 3$ & 2LDB & $-0.024 ^{0.001} _{0.001}$ & $-0.347 ^{0.006} _{0.006}$\\
\hline   
\end{tabular}
\label{Parameters_all}
\end{table}

\begin{figure*}
\begin{center}
\includegraphics[width=3.2in]{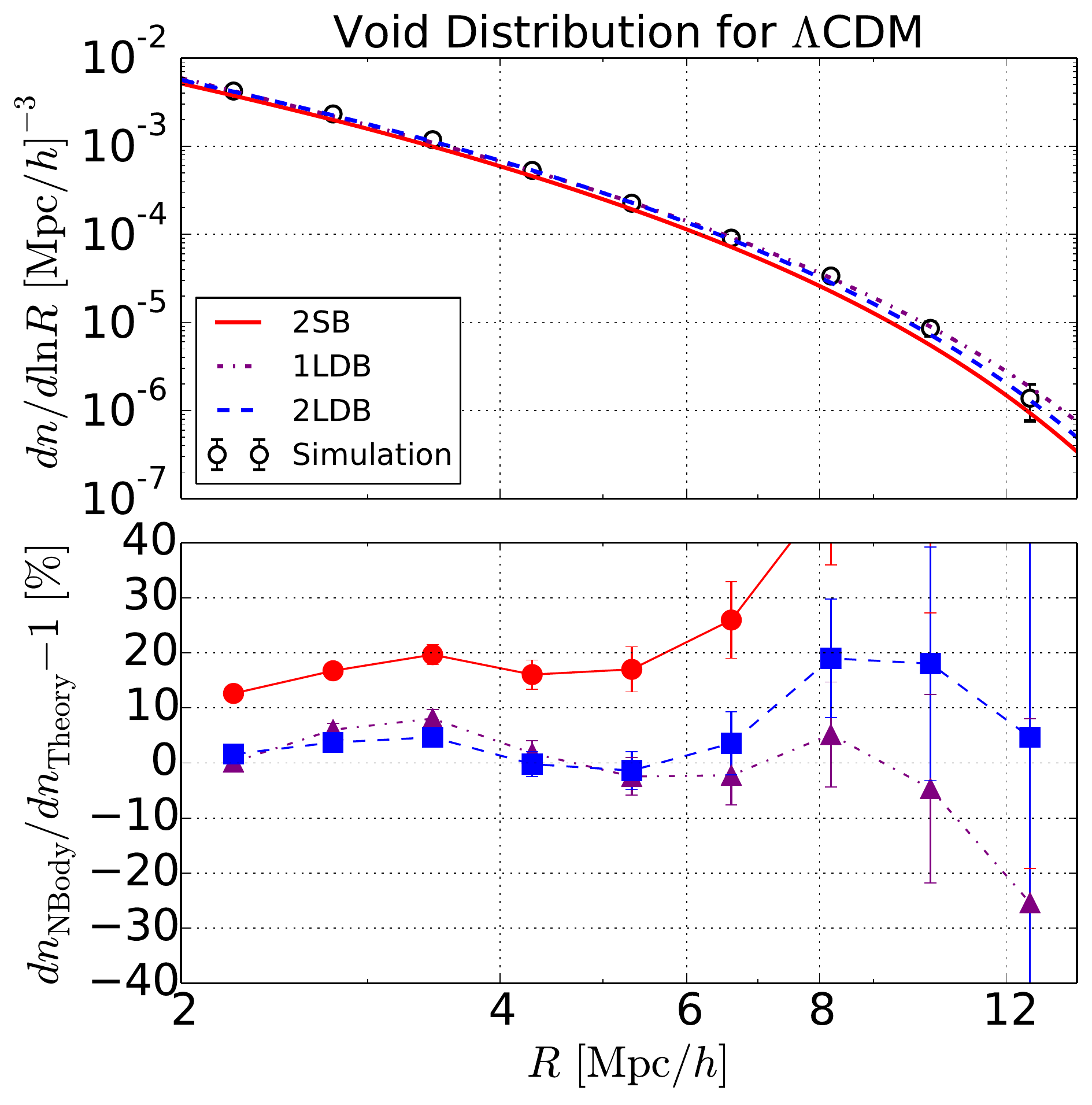}
\includegraphics[width=3.2in]{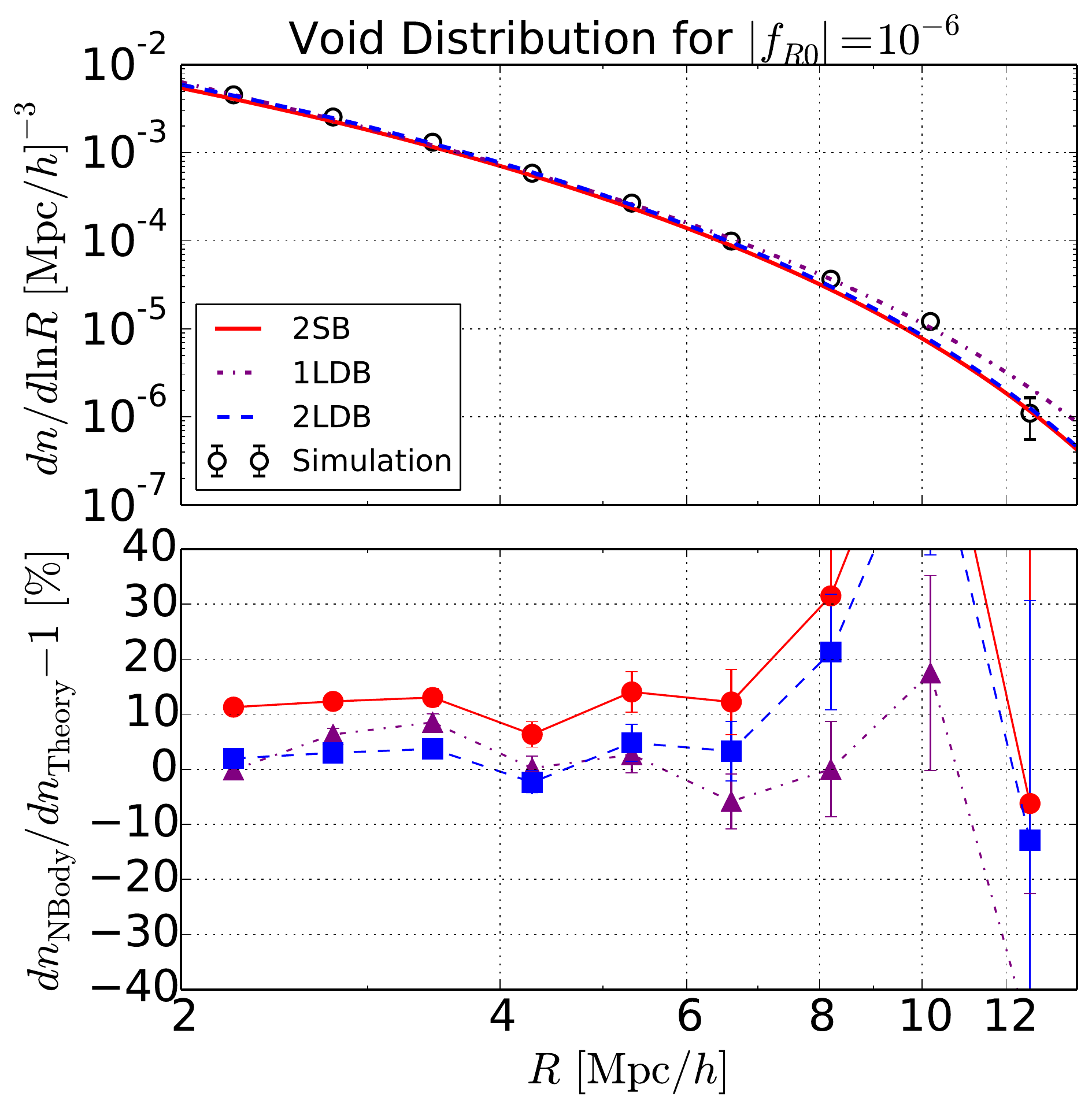}
\includegraphics[width=3.2in]{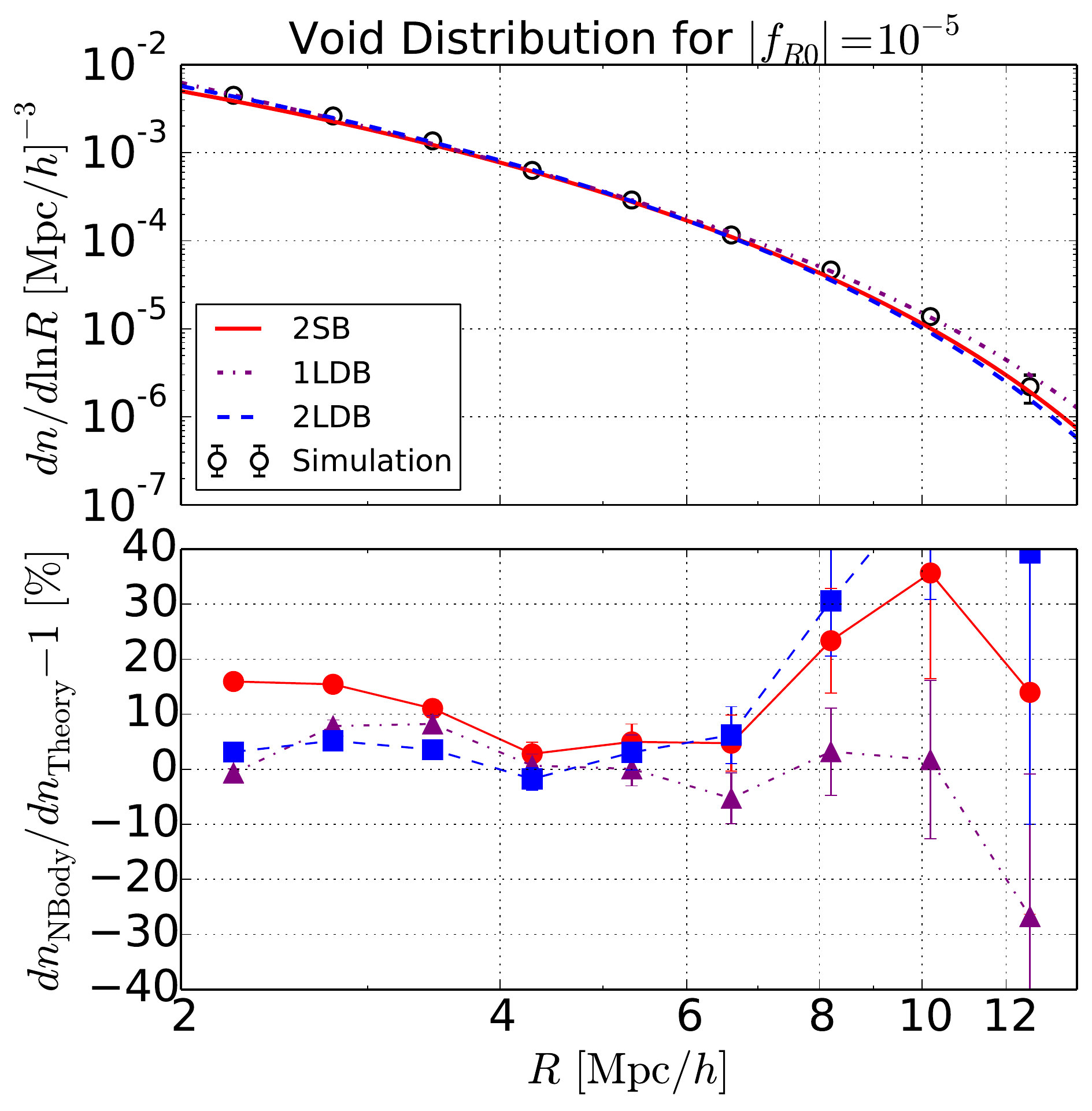}
\includegraphics[width=3.2in]{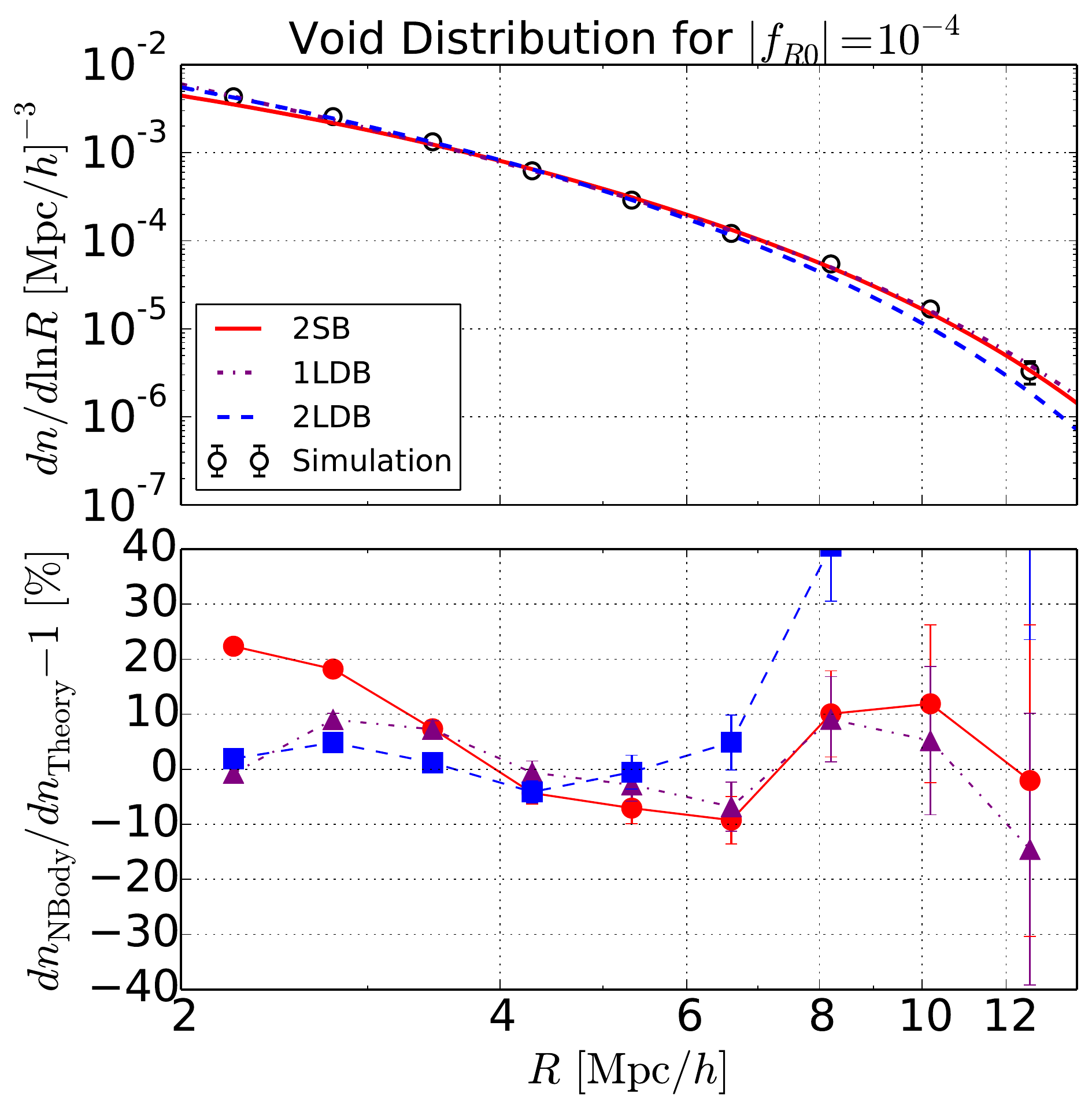}
\caption{ ({\it Top Left}): The upper sub-panel shows the void differential abundance distribution $dn/d\ln R$ as a function of void radius $R$ for GR ($\Lambda$CDM) from simulations ({\it open dots}), along with theory predictions from the 2SB model \cite{Jennings} (red solid curve), from the 1LDB Eq.~\eqref{One_barier} (purple dotted-dashed curve) and the 2LDB model Eq.~\eqref{my} (blue dashed line). The lower sub-panel shows the relative difference between simulation data and 
each theory model. 
 ({\it Top Right}): Same for $f(R)$ modified gravity with $|f_{R0}| = 10^{-6}$.
 ({\it Bottom Left}): Same for $|f_{R0}| = 10^{-5}$
 ({\it Bottom Right}):  Same for $|f_{R0}| = 10^{-4}$
}
\label{Distribution_fofr}
\end{center}
\end{figure*}

\begin{figure*}
\begin{center}
\includegraphics[width=3.2in]{Distribution_lcdm.pdf}
\includegraphics[width=3.2in]{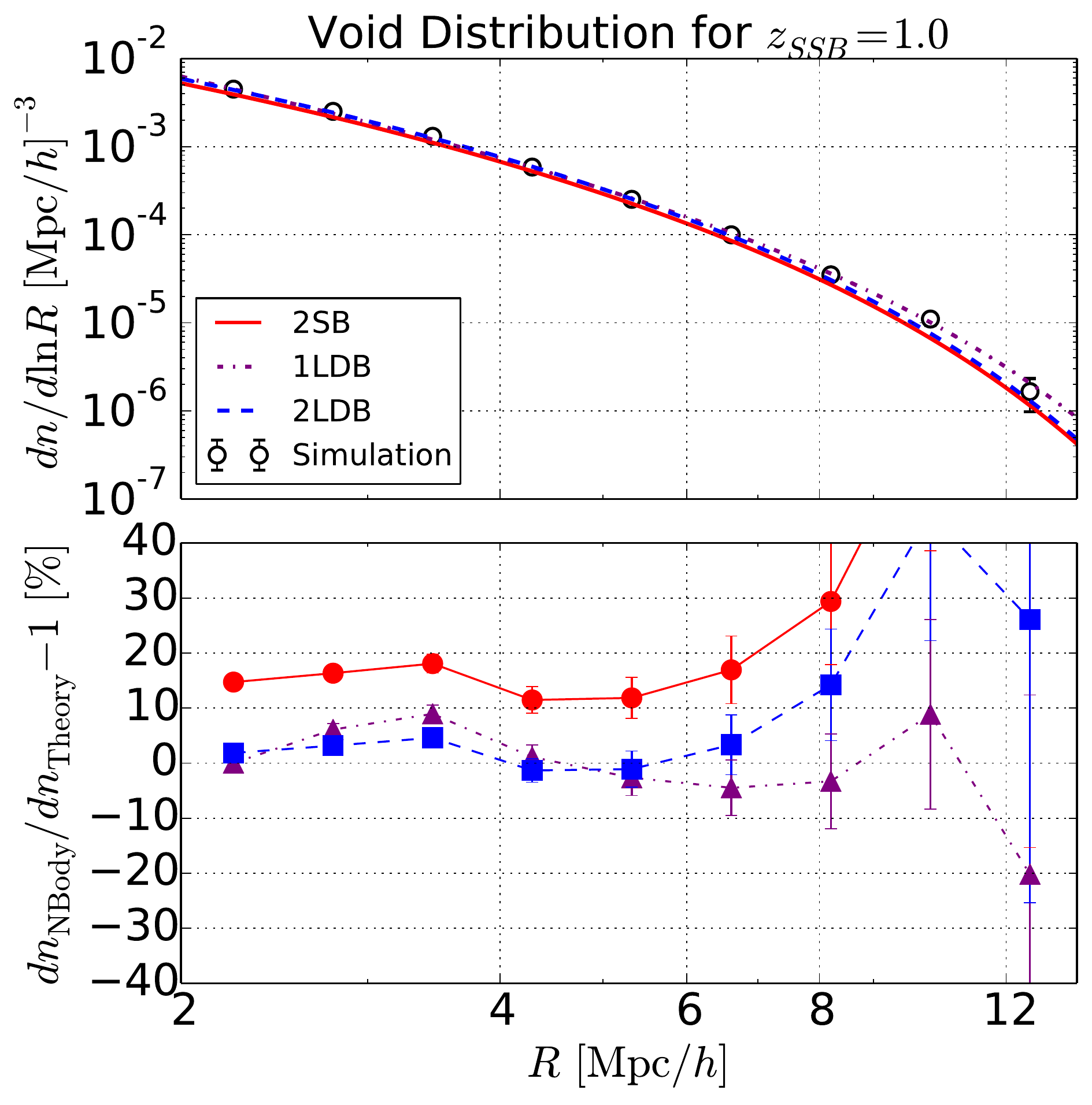}
\includegraphics[width=3.2in]{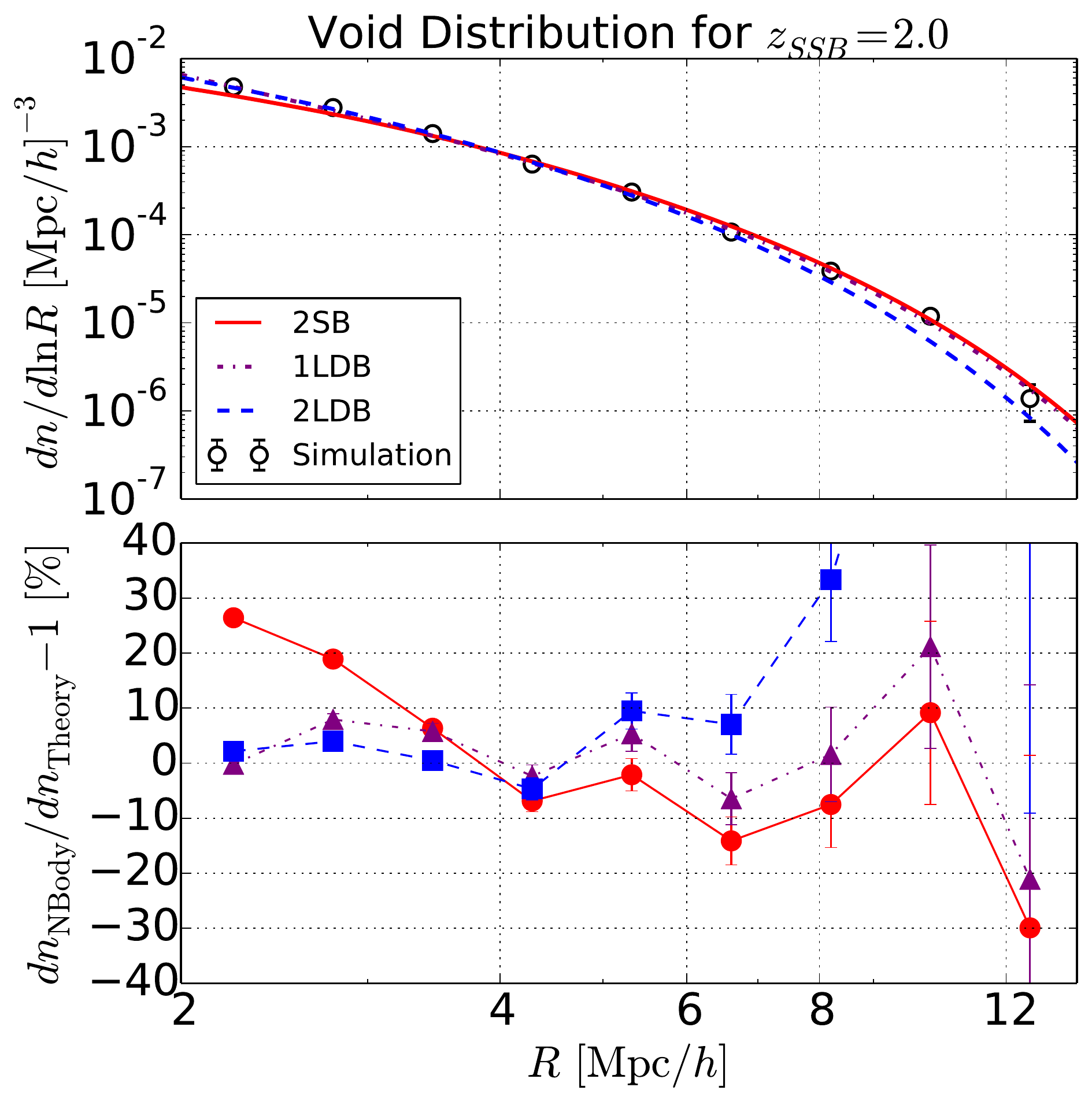}
\includegraphics[width=3.2in]{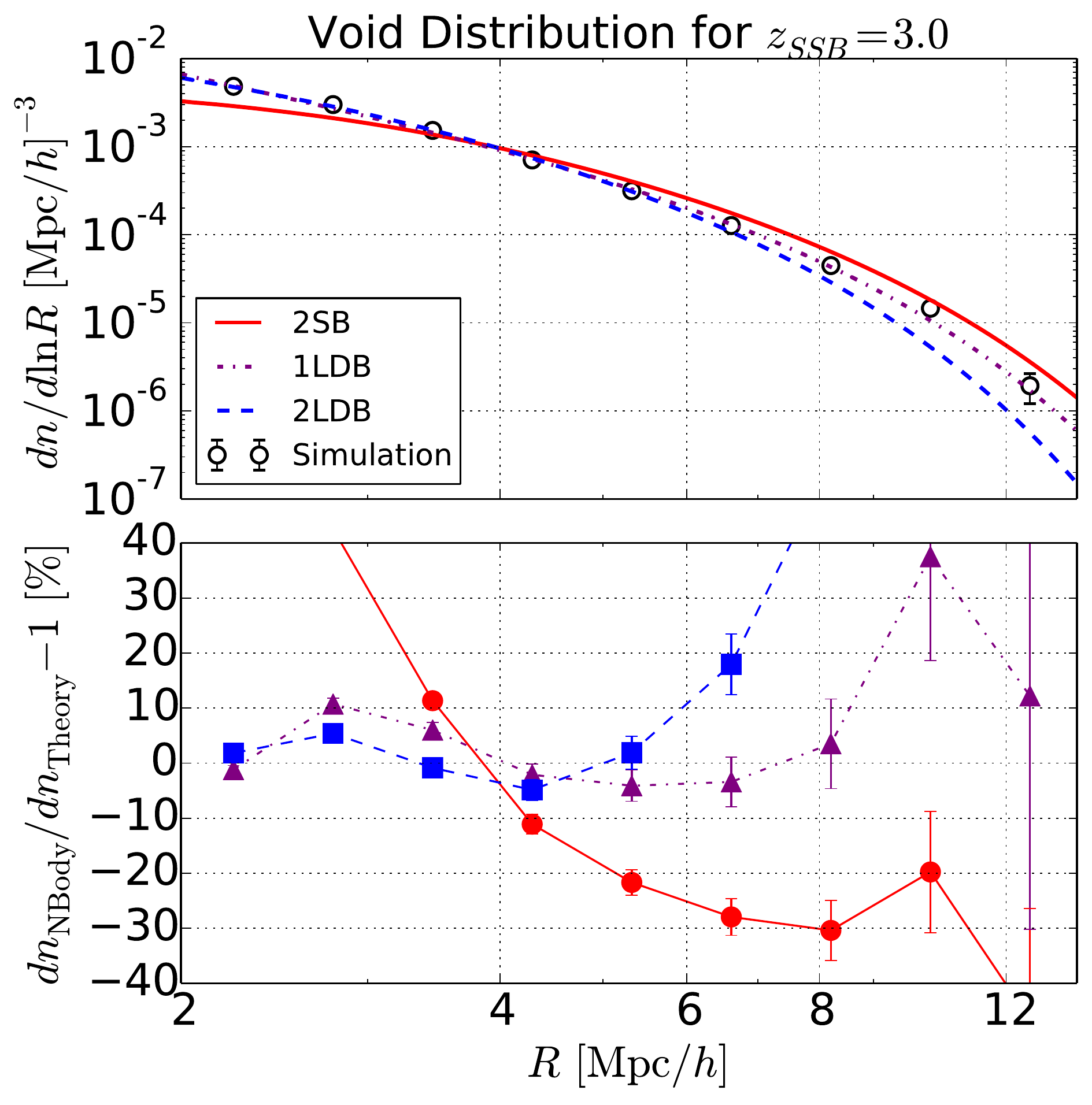}
\caption{Same as Fig.~\ref{Distribution_fofr}, but for the symmetron model with $z_{SSB} = 1$  ({\it top right}), 
 $2$  ({\it bottom left}) and $3$  ({\it bottom right}). 
 }
\label{Distribution_symm}
\end{center}
\end{figure*}

	In Fig.~\ref{Distribution_fofr} we show the abundance of voids $dn/d\ln R$ as measured from simulations (open dots), as well as three theoretical 
	models, namely the 2SB\cite{Jennings}, 1LDB Eq.~\eqref{One_barier} and 2LDB  Eq.~\eqref{my}
	models. 
	Multiple panels show results for $\Lambda$CDM and $f(R)$ models. 
	In Fig.~\ref{Distribution_symm} we show the same for  
	$\Lambda$CDM and symmetron models.  

	We can see that linear-diffusive-barrier models (1LDB and 2LDB) work best in all gravities, relative to the static 
	barriers model (2SB).  In fact, these two models describe the void abundance distribution within $10\%$ precision for 
	$R\lesssim 10$ Mpc/$h$.	
	As expected, the model with two linear diffusive barriers (2LDB) better describes the abundance 
	of small voids  ($R\lesssim 3 $ Mpc/$h$), due to the void-in-cloud effect, more relevant for 
	small voids \cite{Sheth1}.

	In Table~\ref{chi2} we show the reduced $\chi ^{2}$ for GR, the three $f(R)$ models and 
	three symmetron models, 
	This shows again that models with linear diffusive barriers provide a better fit to the simulation data -- 
	with $\chi^2$ one order of magnitude smaller -- and that the 2LDB model gives the 
	overall best fits.
	Another interesting feature for the main model presented 
	in this work (2LDB) is that its reduced $\chi^2$ grows with the intensity of modified gravity. This may indicate 
	that, despite being the best model considered, it may not capture all important features in modified gravity 
	at all orders. We also find that the $f(R)$ model is better fitted than the symmetron model. Since the linear treatment 
	is the same for both gravity models, the 2LDB model may be more appropriate to describe the chameleon 
	screening of $f(R)$ than symmetron screening. Nonetheless,  
	the 2LDB model provides a reasonable representation of the data from both gravity theories in the range considered
	here.

\begin{table}
\centering
\caption{Reduced $\chi ^{2}$ for each gravity model and for the 
three models of void abundance considered.}
\begin{tabular}{c|c|c|c}
\hline 
Gravity & 2SB & 1LDB & 2LDB  \\ 
\hline                           
GR & 15.76 & 3.45 & 1.59 \\ 
$|f_{R0}|=10^{-6}$ & 13.10 & 3.97 & 1.67 \\ 
$|f_{R0}|=10^{-5}$ & 21.10 & 5.52 & 2.11 \\ 
$|f_{R0}|=10^{-4}$ & 34.86 & 5.66 & 2.78 \\ 
$z_{SSB}=1$ & 22.20 & 3.64 & 1.12 \\
$z_{SSB}=2$ & 49.06 & 4.75 & 2.57 \\
$z_{SSB}=3$ & 209.05 & 8.10 &4.77 \\
\hline   
\end{tabular}
\label{chi2}
\end{table} 
	
	As both parameters $\beta$ and $D$ have an explicit dependence on the modified gravity strength, 
	next we fit 
	a relationship between the abundance parameters $\beta$ and $D$ and the gravity 
	parameters $\log _{10} |f_{R0}|$  and $z_{SSB}$.  
	In these fits we set the value $\log _{10} |f_{R0}| = -8$ to represent the case of 
	$\Lambda$CDM cosmology, as this is indeed nearly identical to $\Lambda$CDM for purposes 
	of large-scale structure observables, i.e. $\log _{10} |f_{R0}| = -8 \simeq - \infty $.
	
\begin{figure*}
\begin{center}
\includegraphics[width=1.7in]{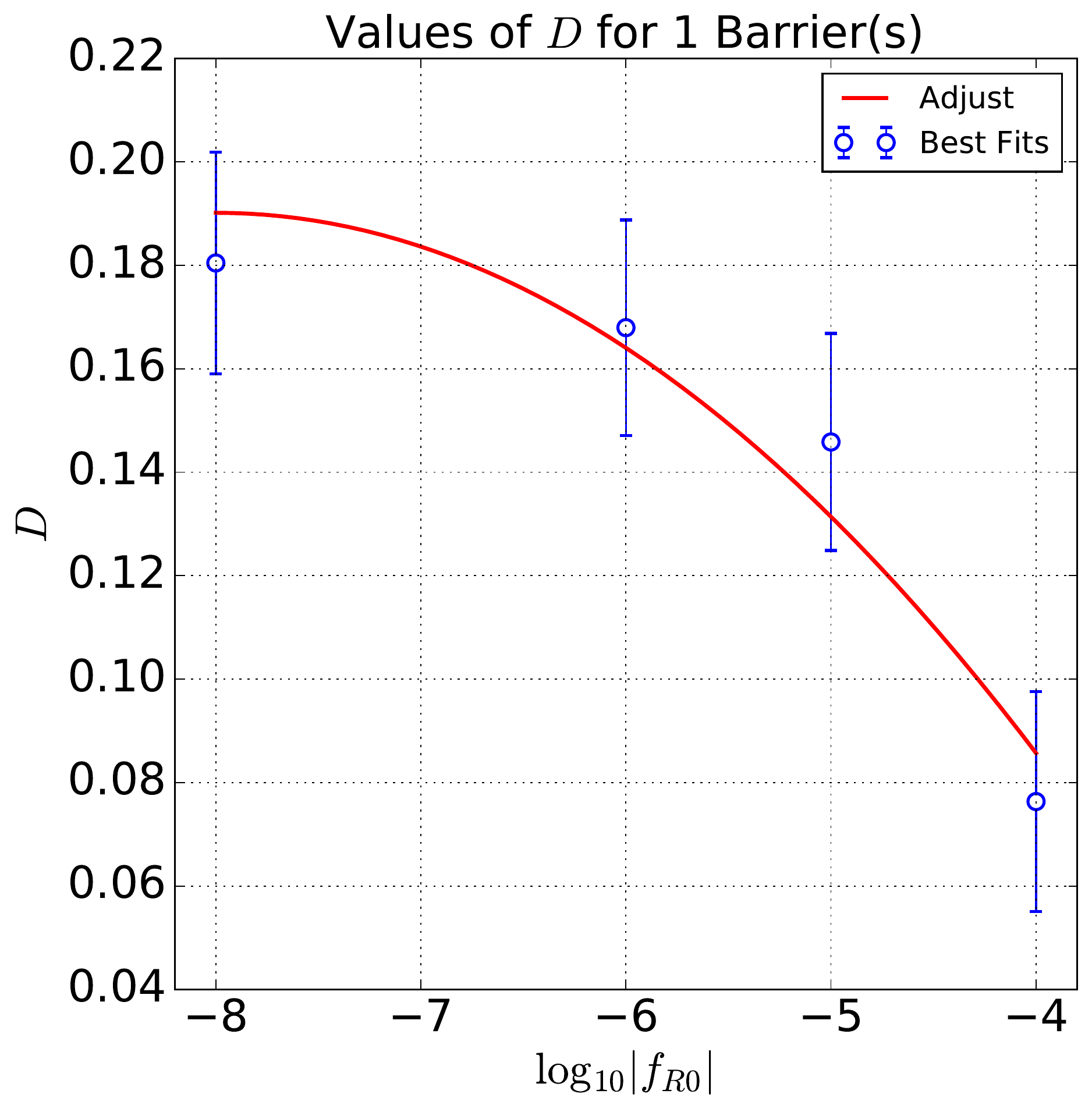}
\includegraphics[width=1.7in]{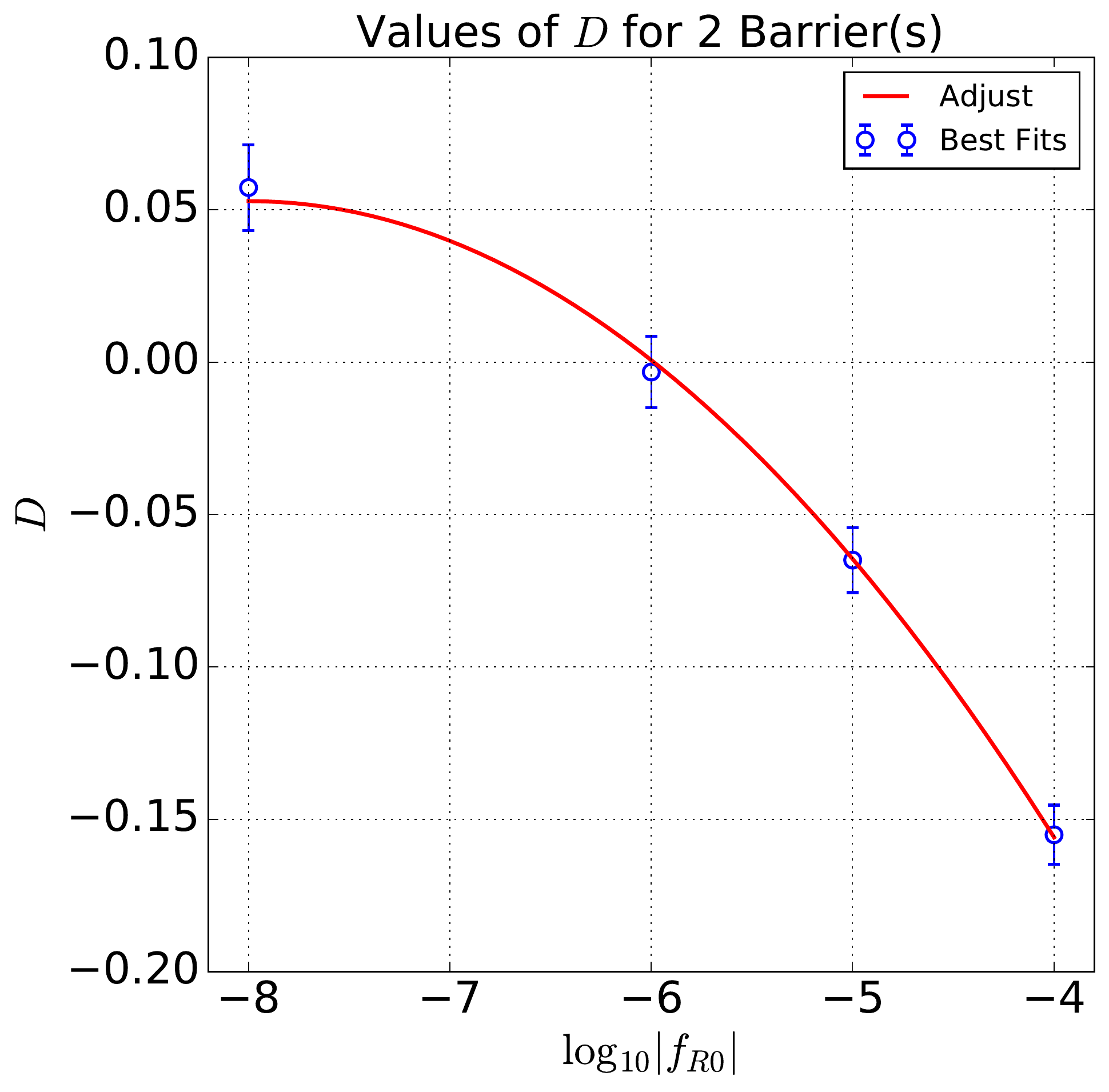}
\includegraphics[width=1.7in]{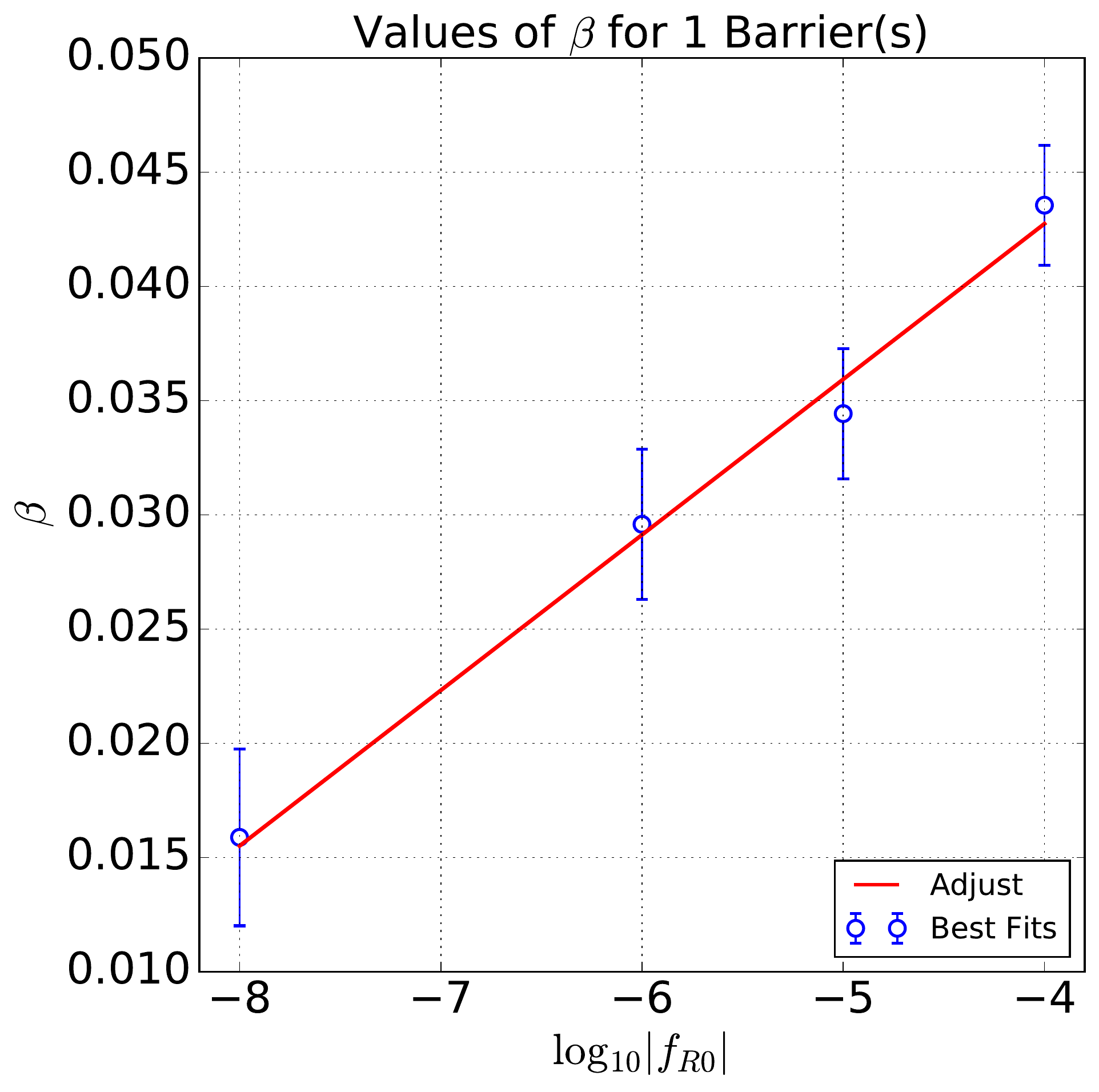}
\includegraphics[width=1.7in]{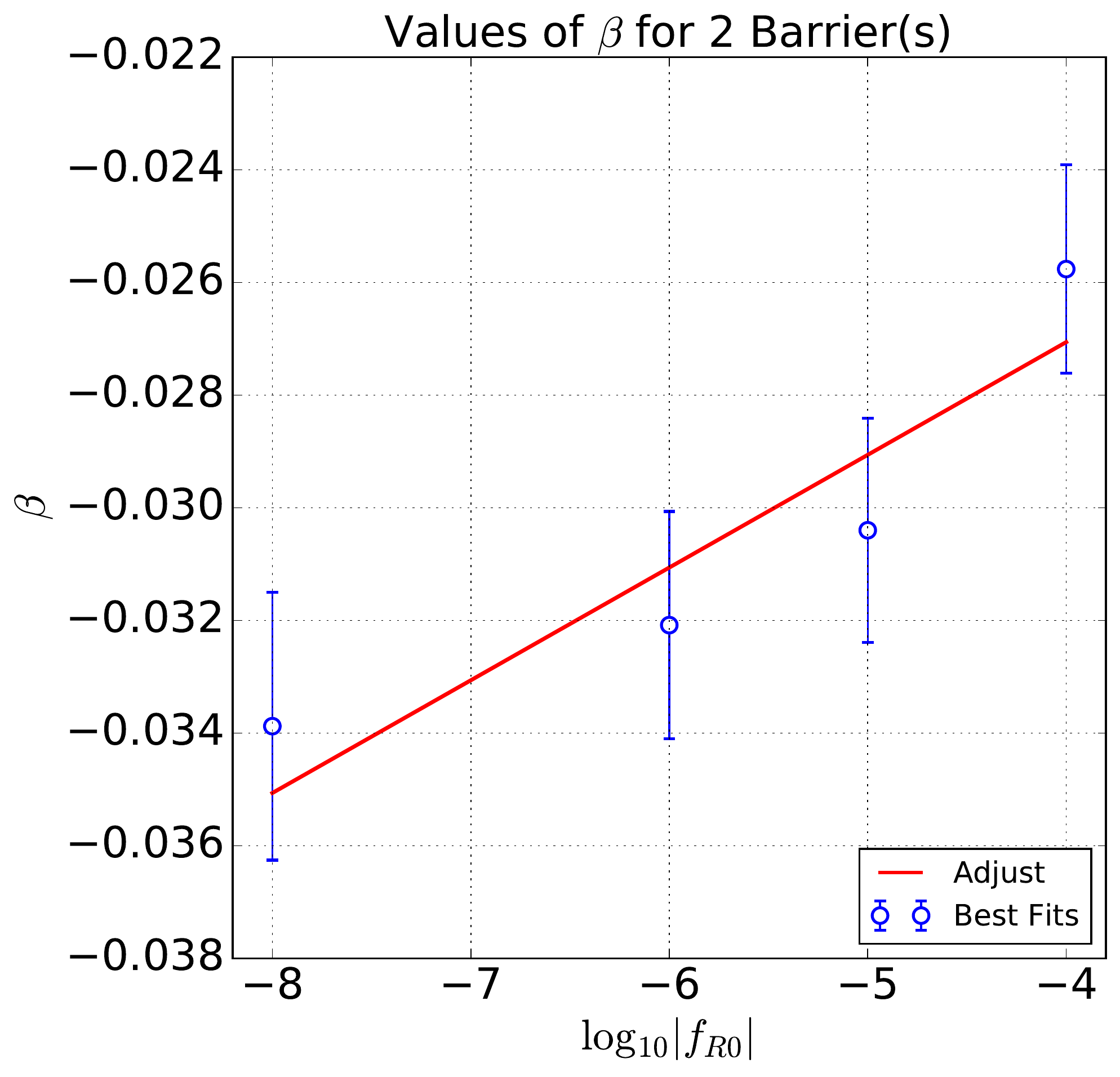}
\includegraphics[width=1.7in]{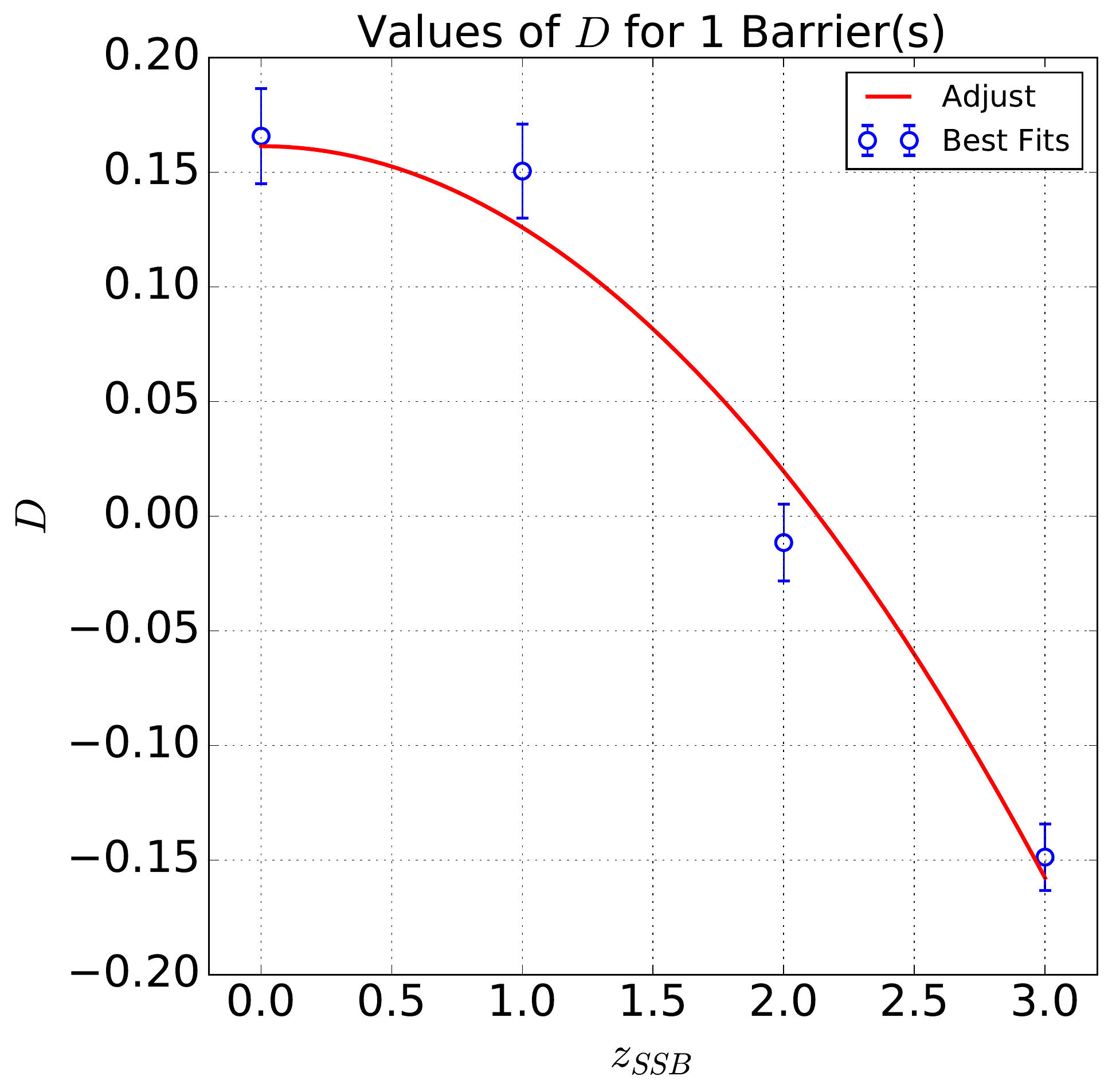}
\includegraphics[width=1.7in]{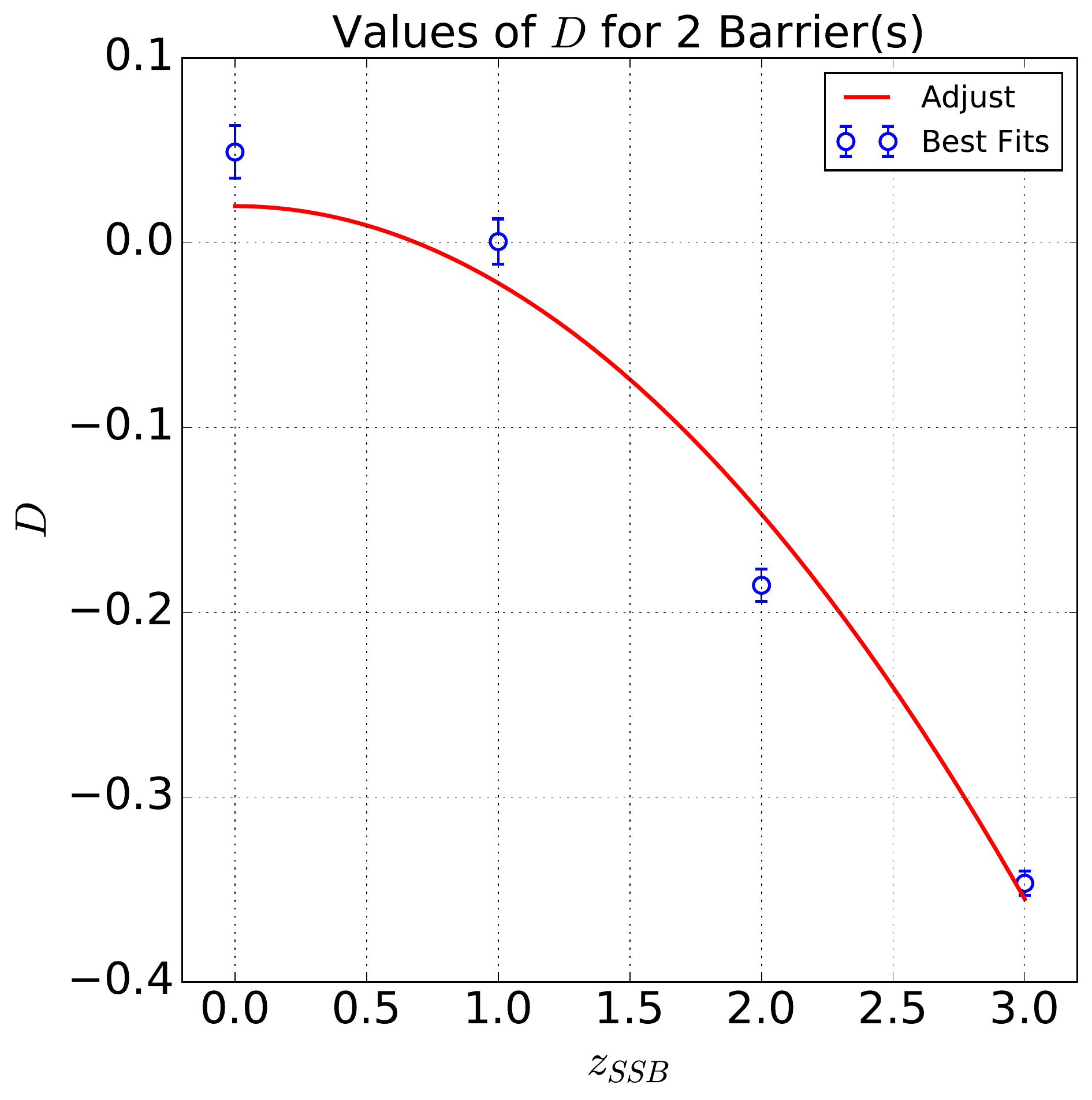}
\includegraphics[width=1.7in]{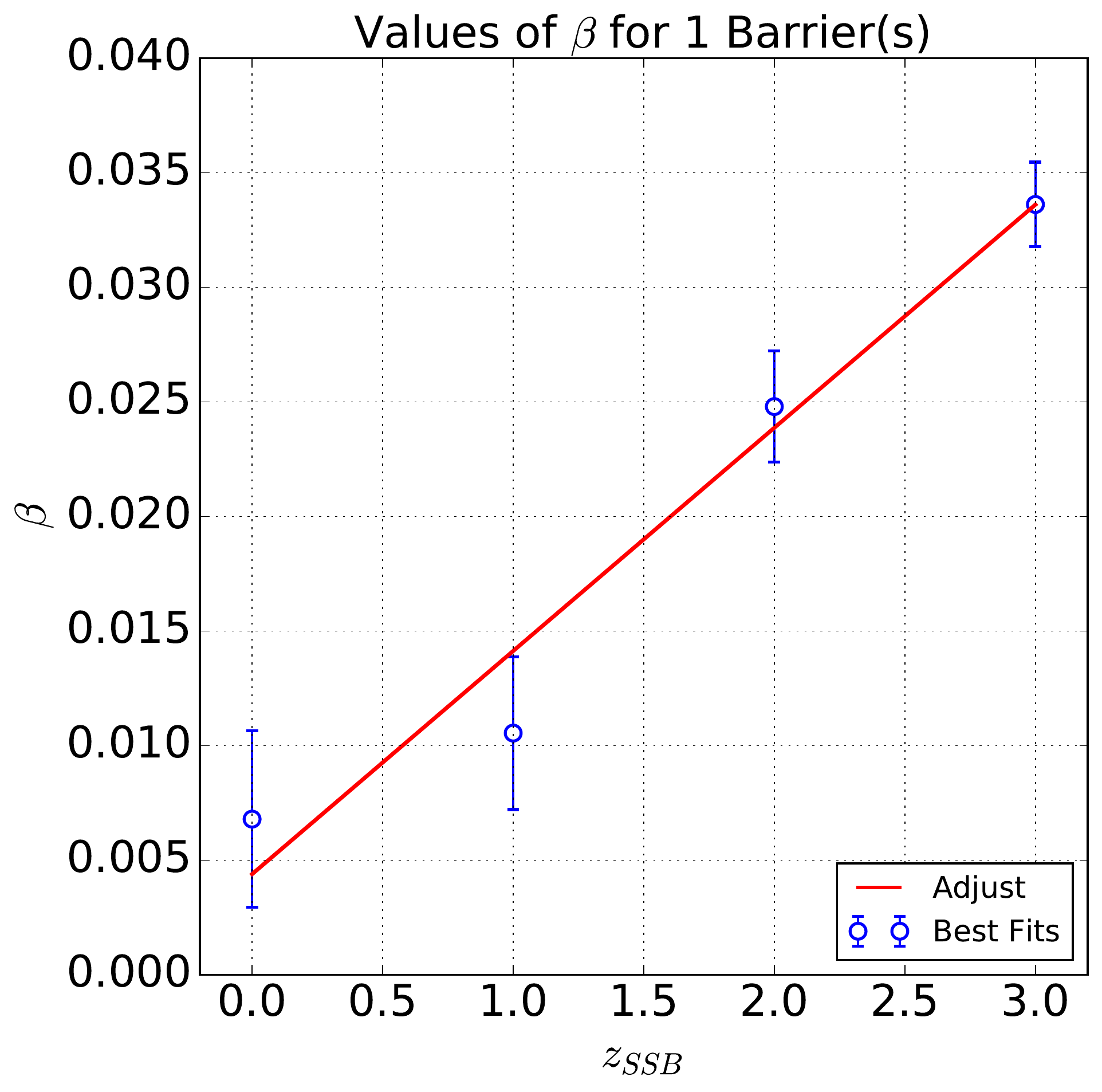}
\includegraphics[width=1.7in]{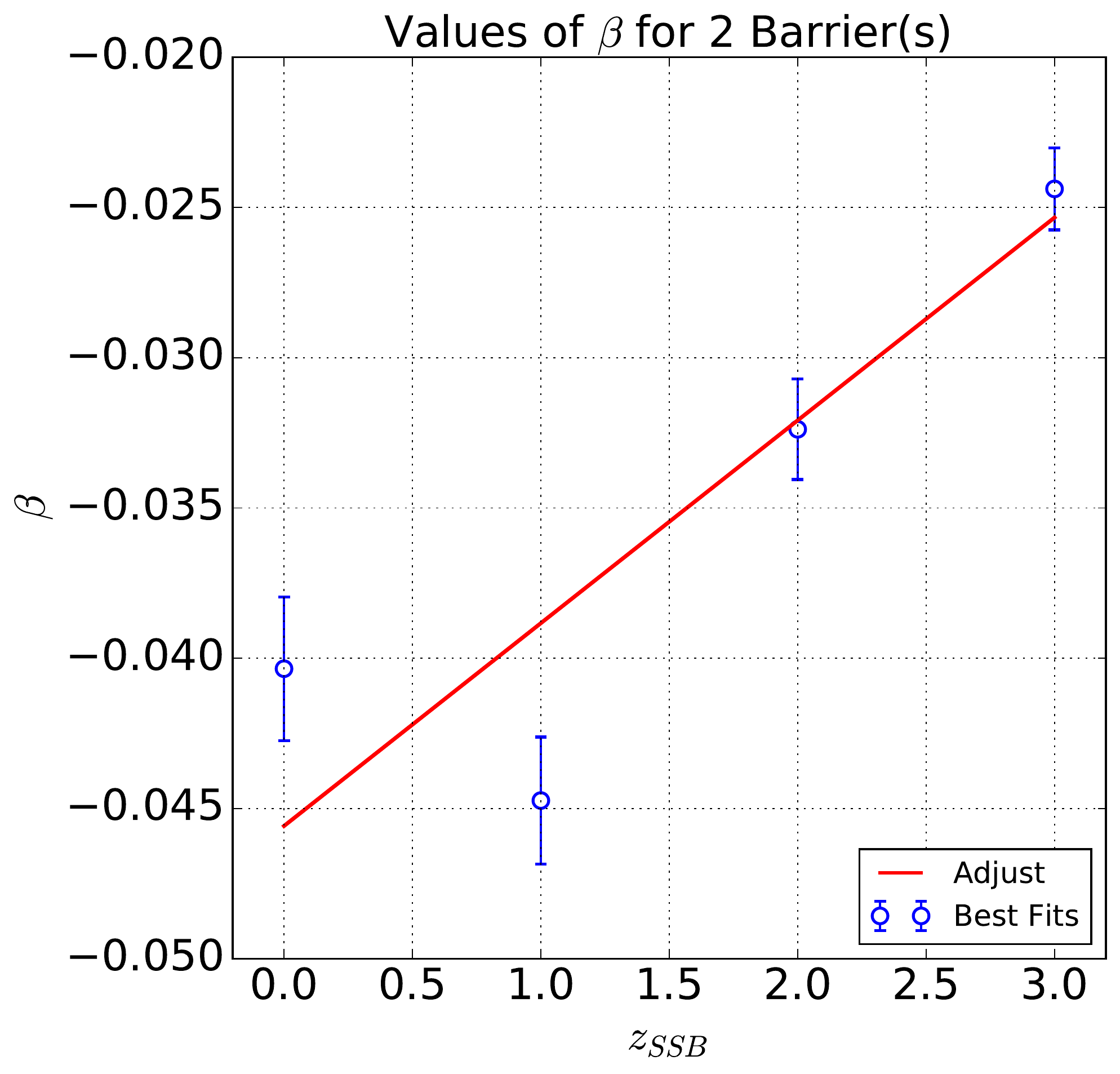}
\caption{ ({\it Top Row}): Fits of $D$ and $\beta$ as a function of $\log_{10}|f_{R0}|$ in $f(R)$ gravity. These fits are shown for $D$ in the 1LDB and 2LDB models, 
and for $\beta$ in the 1LDB and 2LDB models respectively from left to right.
 ({\it Bottom Row}): Same for fits as a function of $z_{SSB}$ in symmetron gravity.
}
\label{Param_fits}
\end{center}
\end{figure*}

        As we expect $\beta$ and $D$ to depend monotonically on the modified gravity 
        parameters, we fit for them using simple 
        two-parameter functions. For $\beta$ case we use a straight line, and for $D$ a 
        second order polynomial with maximum fixed by the $\Lambda$CDM value. 
        These fits are shown in the multiple panels of Fig.~\ref{Param_fits}.
	
	Our values of $\beta$ and $D$ as a function of gravity parameters 
	fluctuate considerably around the best fit. This occurs at least partially because 
	we have used only one simulation for each gravity model, and we expect this oscillation 
	to be reduced with a larger number of simulations. At present, the use of the fits is likely 
	more robust than the use of exact values obtained for each parameter/case.  	
	
\subsection{Constraining Modified Gravity}

	Given the fits for $\beta$ and $D$ obtained in the last subsection, we now check for the power 
	of constraining modified gravity from the void distribution function in each of the three void 
	abundance models considered, namely 2SB, 1LDB and 2LDB. 
	We take the abundance of voids actually found in simulations (described in the \S IV) to 
	represent a hypothetical real 
	measurement of voids and compare it to the model predictions, evaluating the   
	posterior for $\log_{10}|f_{R0}|$ and $z_{SSB}$, 
	thus assessing the constraining power 
	of each abundance model in each gravity theory. 
	Obviously the constraints obtained in this comparison are optimistic -- since we are taking 
	as real data the same simulations used to fit for the abundance model parameters -- but 
	they provide us with idealized constraints 
	similar in spirit to a Fisher analysis around a fiducial model. 
	
\begin{figure*}
\begin{center}
\includegraphics[width=3.2in]{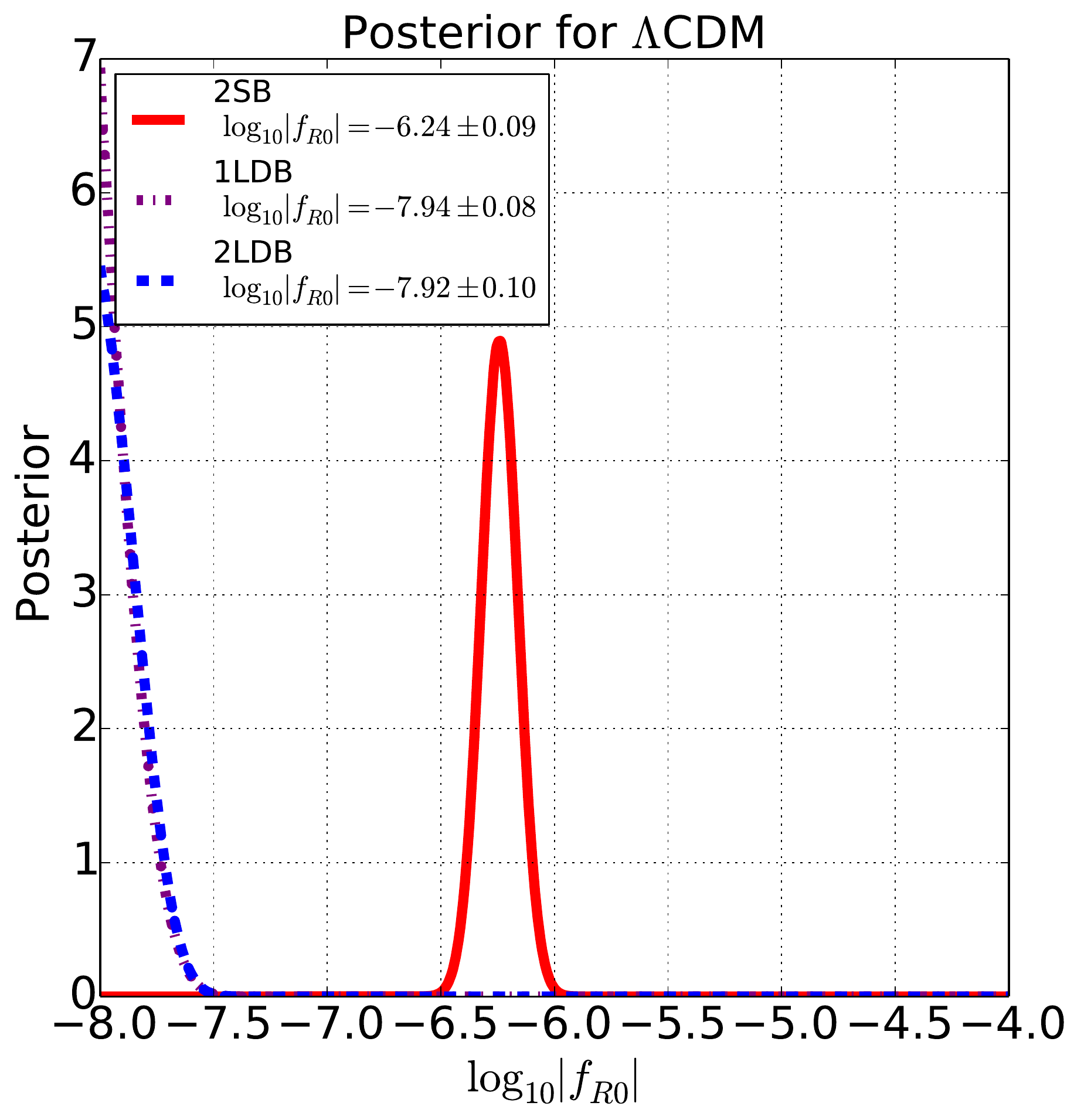}
\includegraphics[width=3.2in]{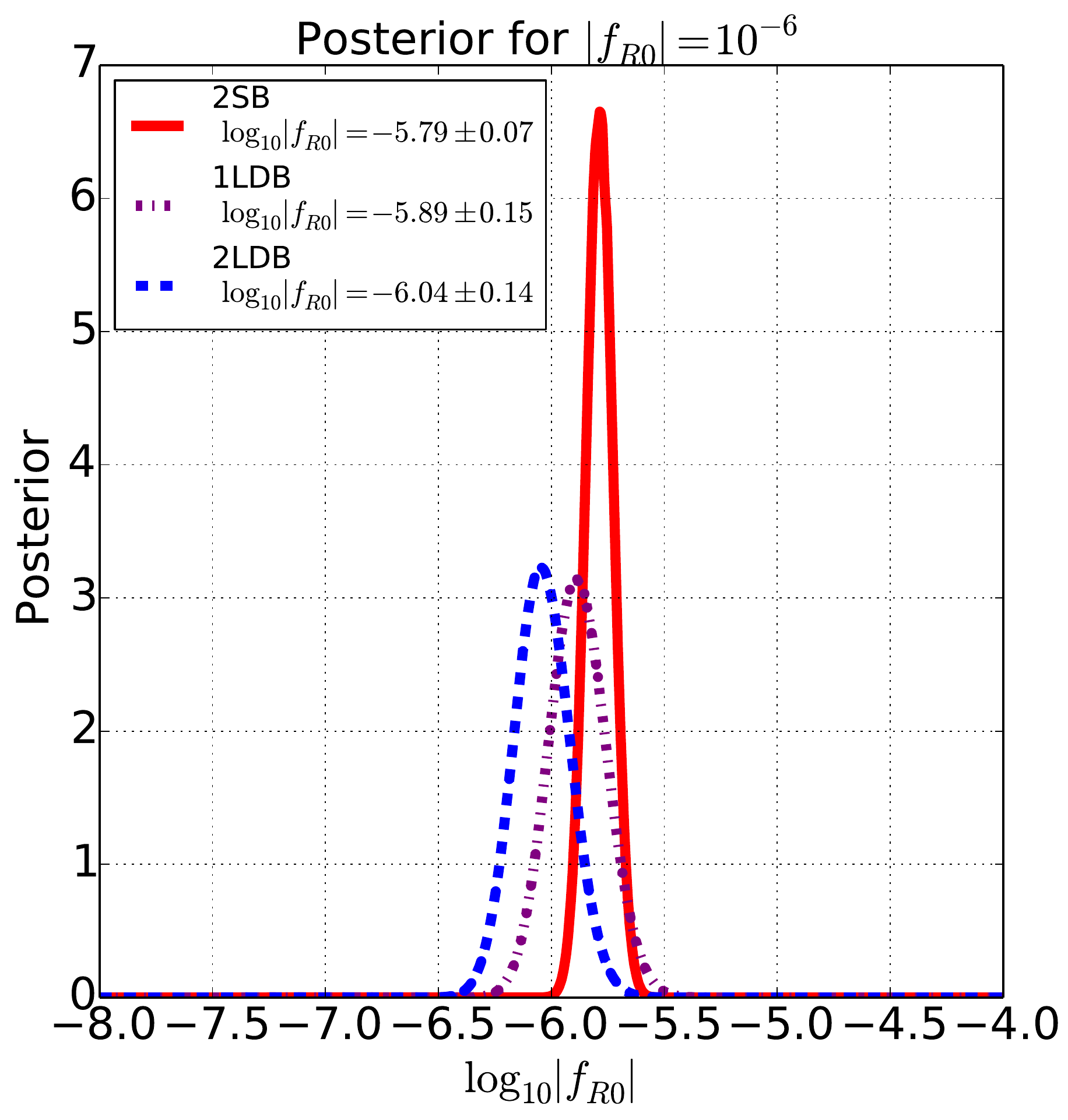}
\includegraphics[width=3.2in]{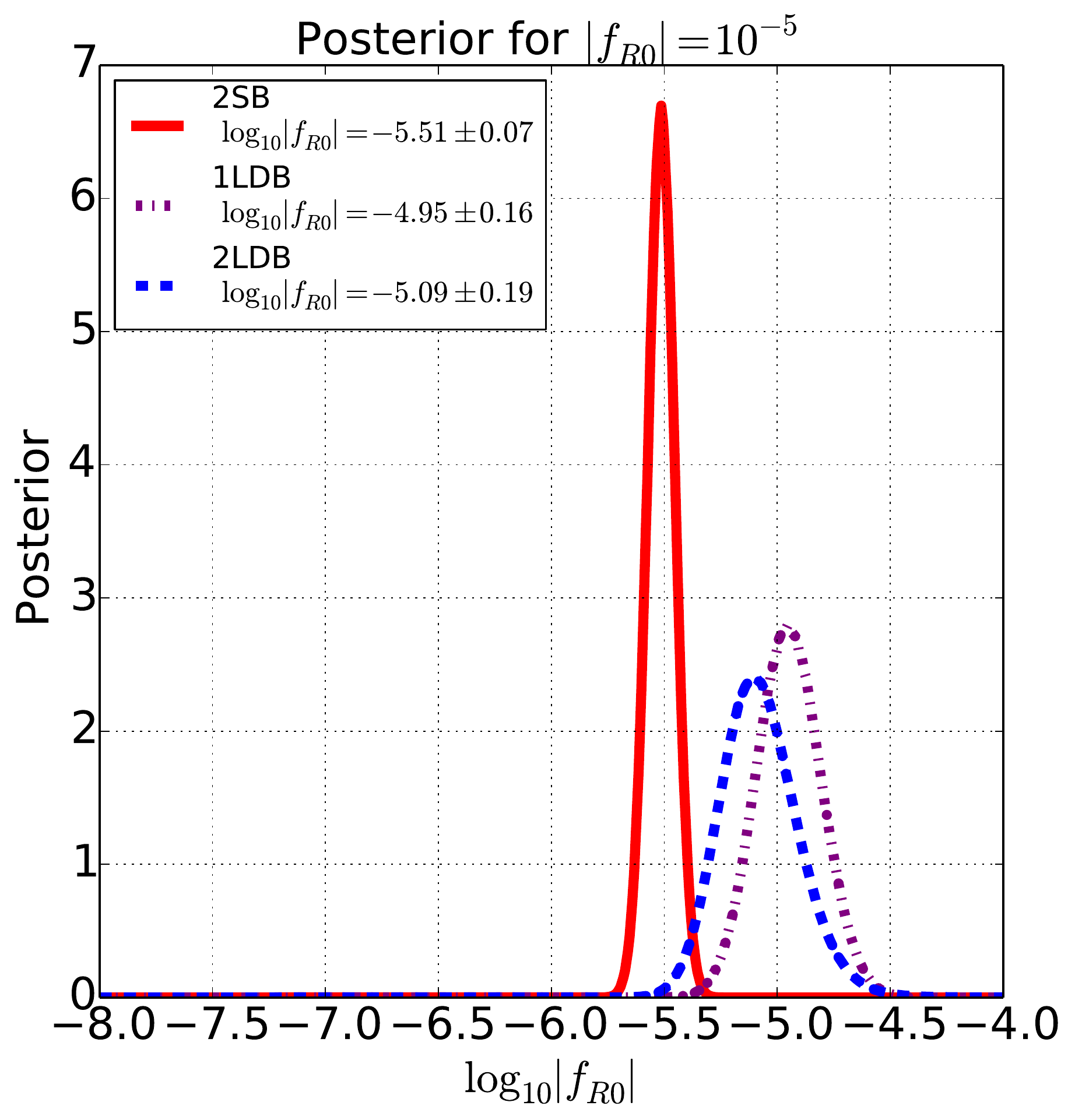}
\includegraphics[width=3.2in]{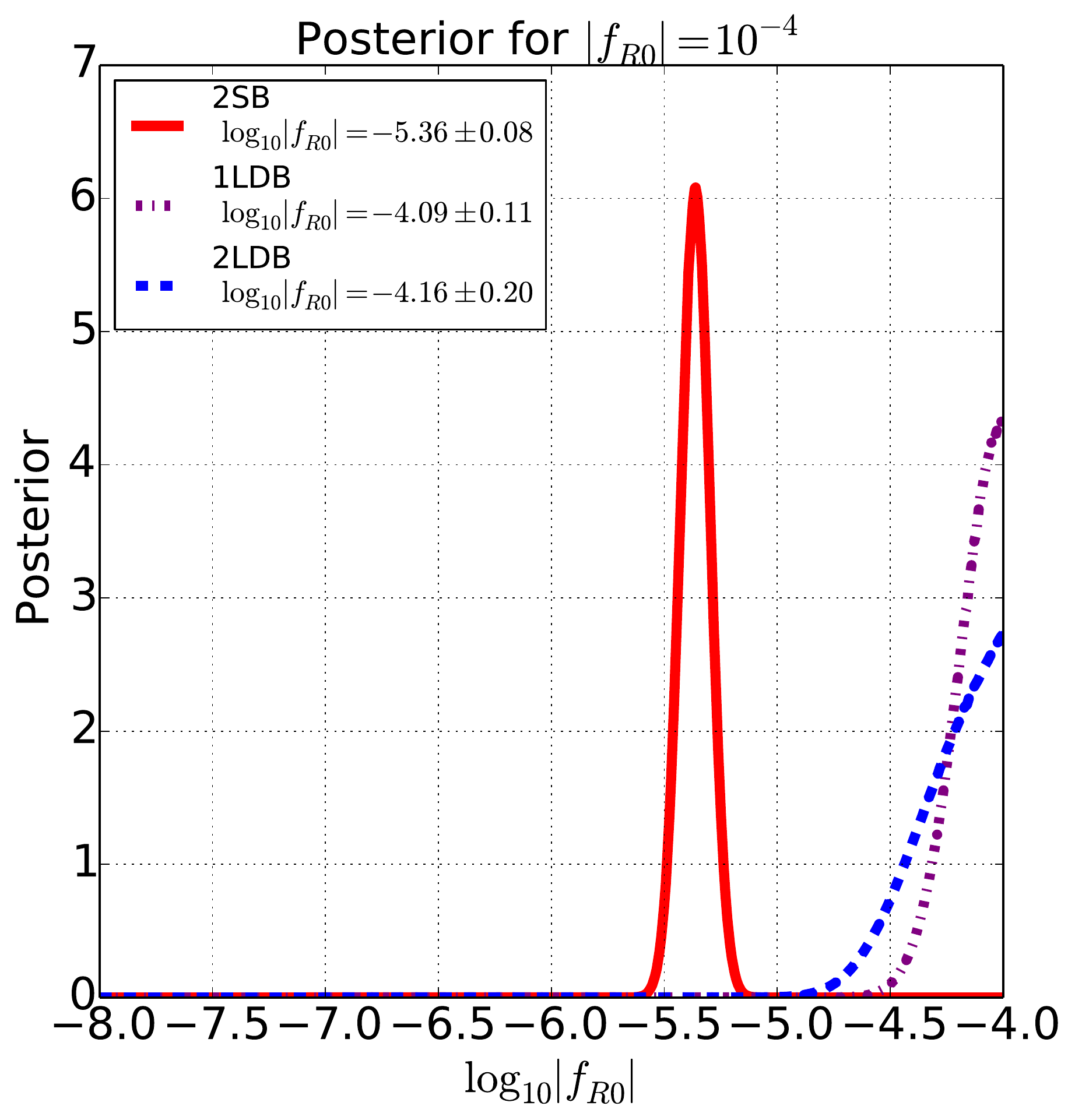}
\caption{Posterior distribution for $\log _{10} |f_{R0}|$ and for the three abundance models considered in the text, 2SB model \cite{Jennings} (red continuous line), 2LDB model Eq.~\eqref{my} (blue dashed line) and 1LDB model Eq.~\eqref{One_barier} (purple dotted dashed line). The mean and 1$\sigma$ values of $\log _{10} |f_{R0}|$ in each case are indicated in the legend. 
({\it Top Left}): Posterior for the $\Lambda$CDM simulation.
({\it Top Right}): Posterior for the $|f_{R0}| = 10^{-6}$.
({\it Bottom Left}): Posterior for the $|f_{R0}| = 10^{-5}$.
({\it Bottom Right}): Posterior for the $|f_{R0}| = 10^{-4}$.
}
\label{Posterior_fofr}
\end{center}
\end{figure*}

\begin{figure*}
\begin{center}
\includegraphics[width=3.2in]{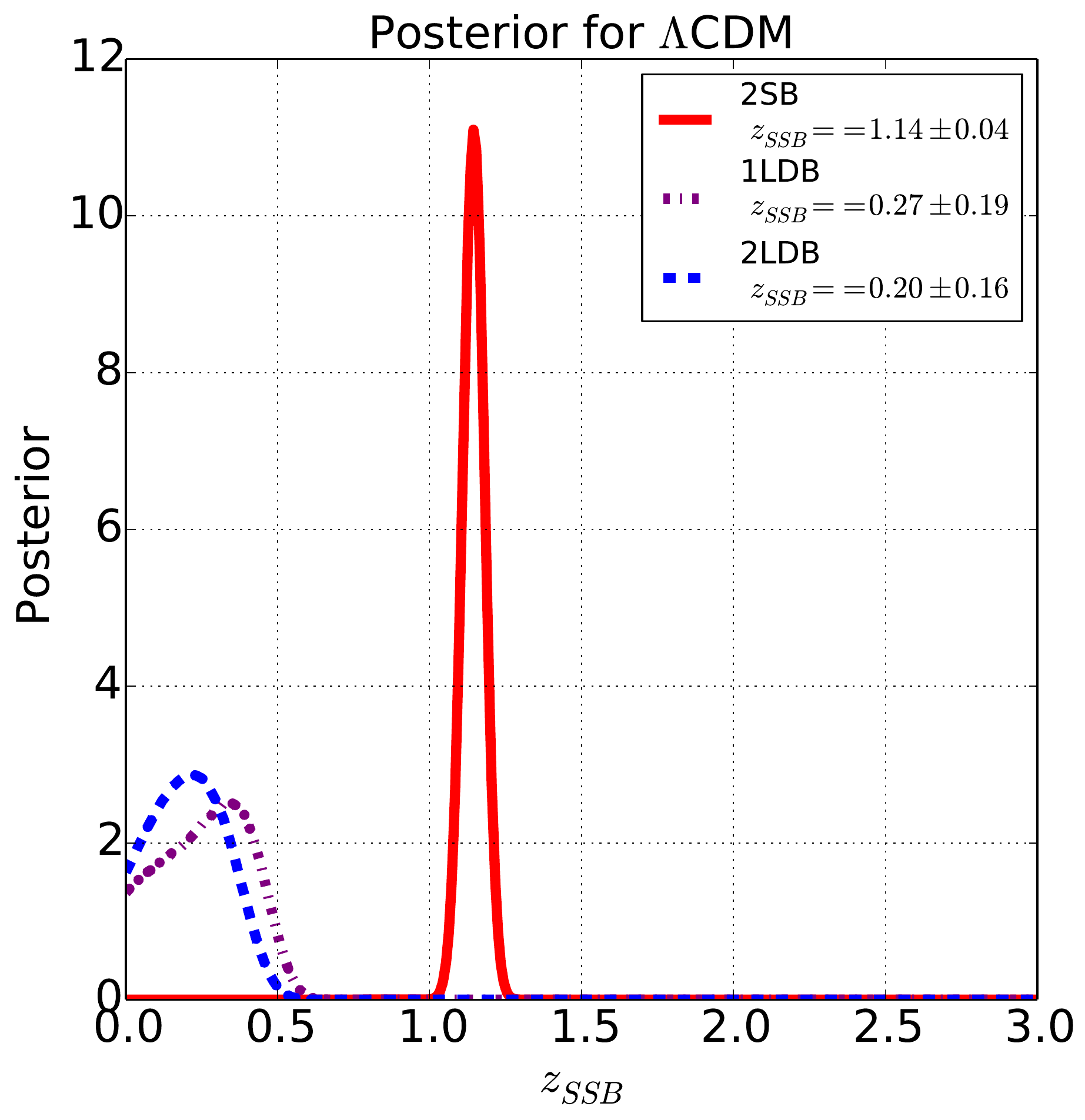}
\includegraphics[width=3.2in]{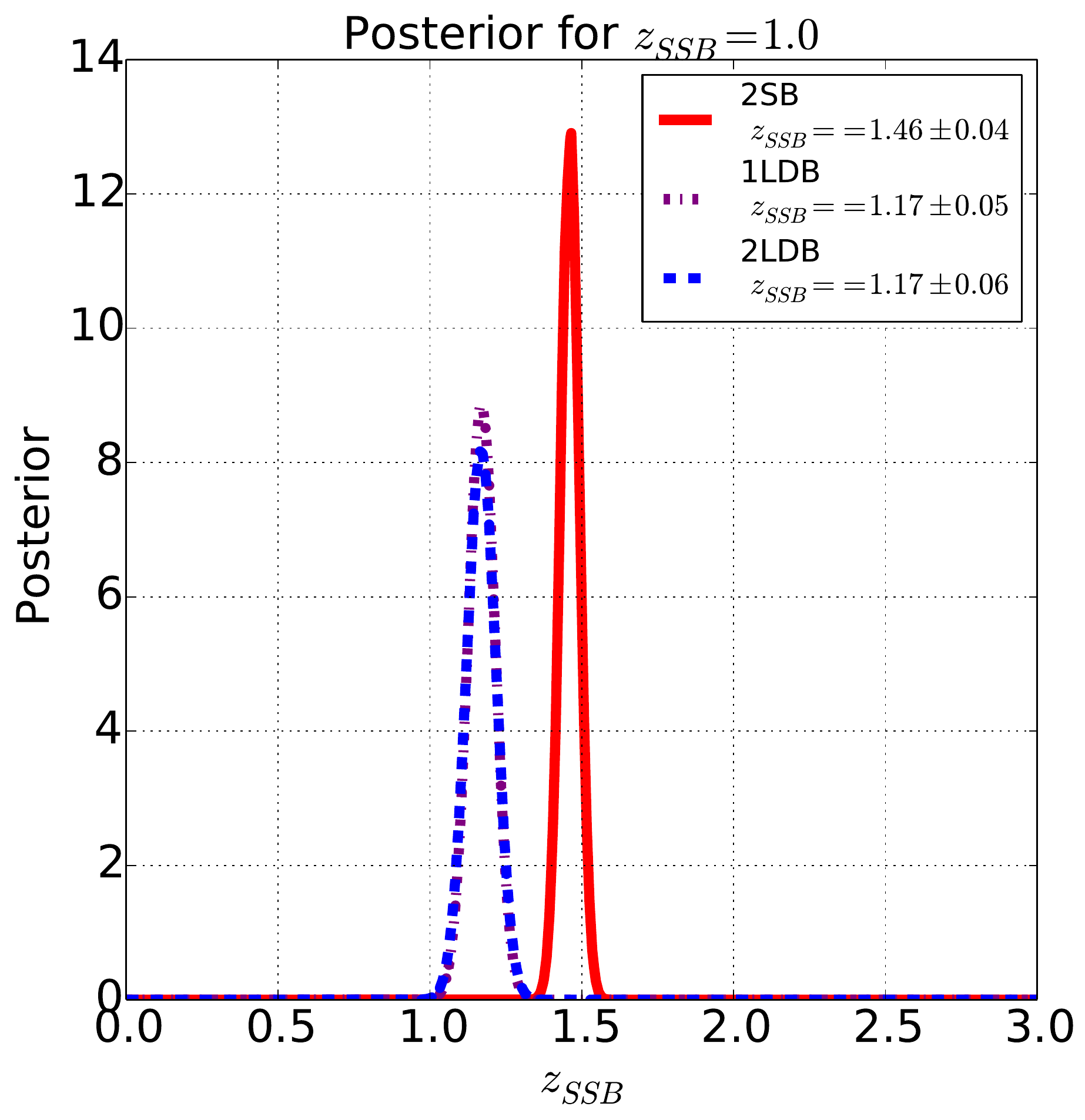}
\includegraphics[width=3.2in]{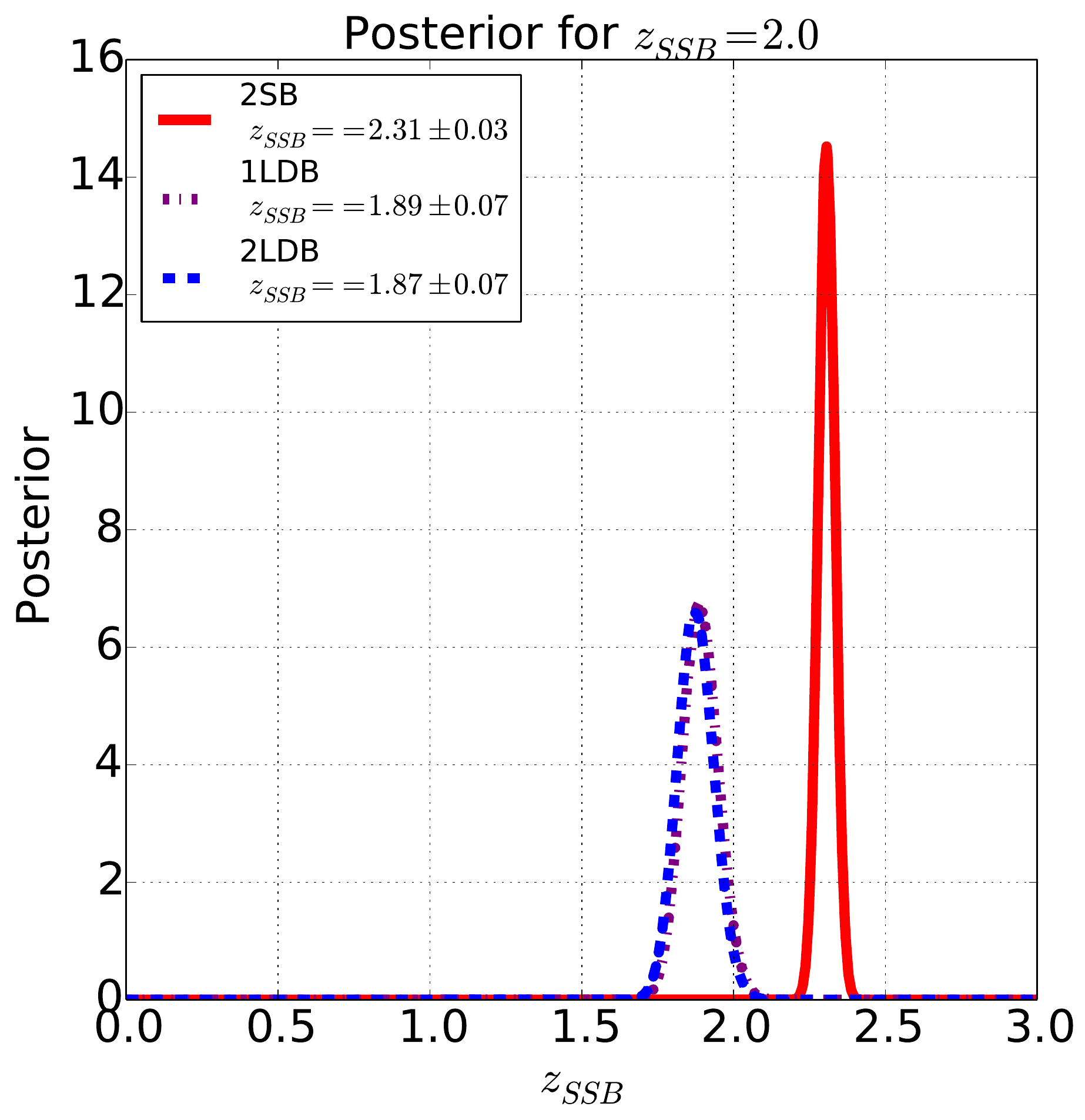}
\includegraphics[width=3.2in]{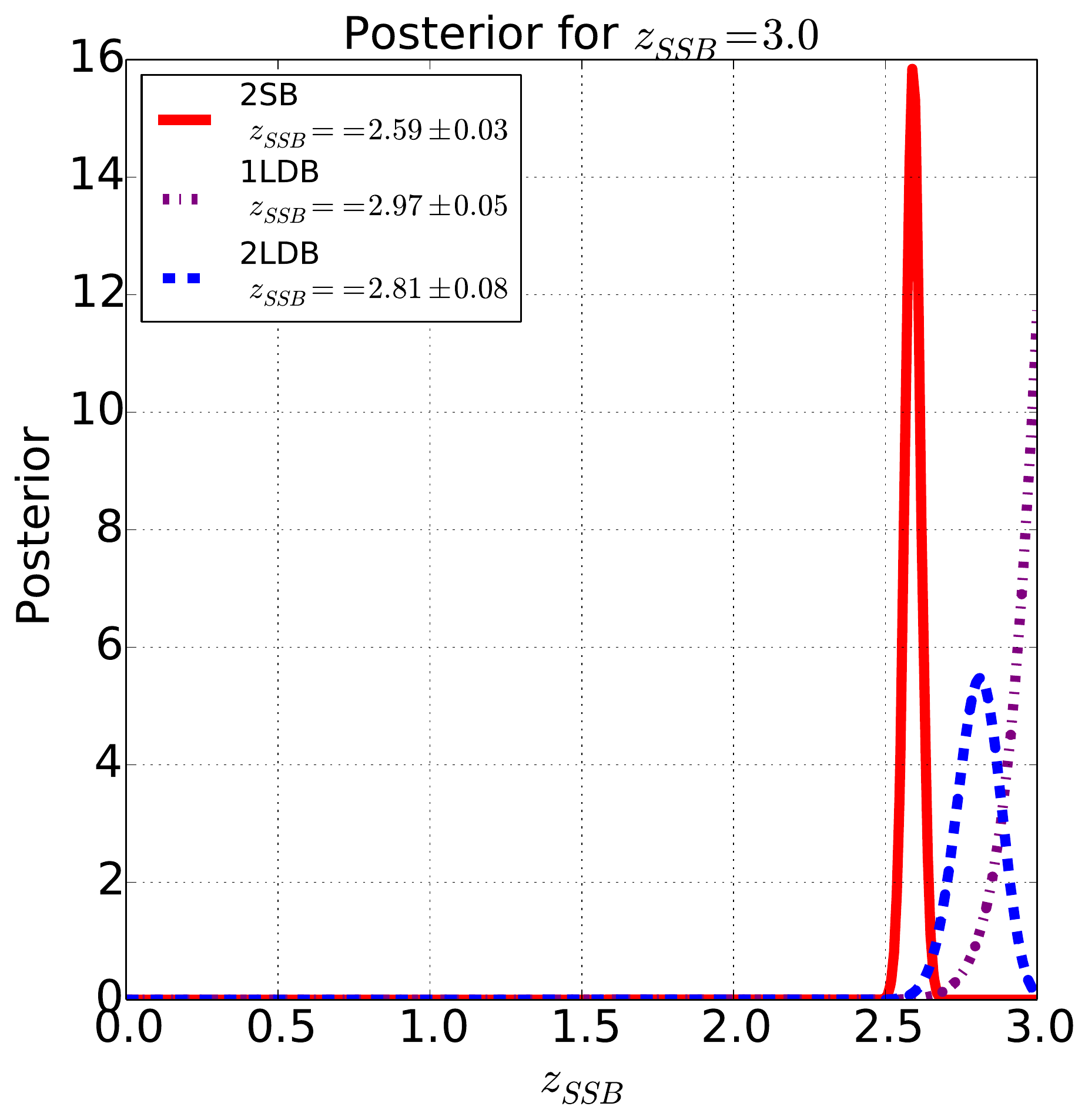}
\caption{Same as Fig.~\ref{Posterior_fofr}, but for the symmetron model with $z_{SSB} = 1$ ({\it top right}), 
2 ({\it bottom left}) and 3 ({\it bottom right}). }
\label{Posterior_symm}
\end{center}
\end{figure*}

	The posteriors for the gravity parameters are shown in Figs.~\ref{Posterior_fofr} and 
	\ref{Posterior_symm}, as well as the 
	mean values and 1$\sigma$ errors in each case. For the results shown here all 
	cosmological parameters 
	 from \S~\ref{sec:void_sim} have been fixed to their true values. We also considered the case where 
	 we apply Planck priors \cite{Planck15} on $\Omega _{dm}$ 
	 and $h$ and let them vary freely in the MCMC, keeping other parameters fixed. 
	 In the latter case,  
	 the mean values and errors found for $\log_{10} |f_{R0}|$ are slightly worse, but the errors remain 
	 less than twice those found for 
	 the case of all fixed parameters.  Moreover, the errors derived for $\Omega_{dm}$ and $h$ 
	 reduce to half of their original Planck priors. 
	
	In Fig.~\ref{Posterior_fofr} we can see that the 2SB model predicts values for the $f(R)$ parameter 
	($\log _{10} |f_{R0}|$) which are incorrect by more than 3$\sigma$ for all cases. In fact, this 
	model predicts incorrect values even for general relativity. This is not surprising given the bad 
	$\chi^2$ fits from Table~IV. Therefore we find this model to be highly inappropriate to describe 
	the abundance of dark matter voids, and focus on 
	models with linear diffusive barriers. 
	
	Both the 1LDB and 2LDB models predict correct values for the gravity parameters within 
	1$\sigma$ in most cases. 
	We find that the 1LDB model presents results similar to 2LDB, despite being a simpler model 
	and providing a worse fit to the data (larger reduced $\chi ^{2}$). 
	For $\Lambda$CDM both posteriors go to 
	$\log _{10} |f_{R0}| = 10^{-8}$, which represents the GR case by assumption.  
	This shows that within the $f(R)$ framework, we can also constrain GR with reasonable precision 
	from void abundance, using  
	one of these two abundance models with diffusive barriers (1LDB, 2LDB).
	
	For the symmetron Model, we can see in Fig.~\ref{Posterior_symm} that the parameter $z_{SSB}$ 
	is also well constrained, similarly to $f_{R0}$ in $f(R)$. 
	Again the 2SB model has the worst result in all cases, and the 1LDB and 2LDB models produce 
	similar results.

	In Table~\ref{Best_and_mean} we show the best-fit values, mean values and 1$\sigma$ errors 
	from the posteriors distributions of Figs.~\ref{Posterior_fofr}, \ref{Posterior_symm} for the $f(R)$ 
	and symmetron theories. It becomes again clear that our proposed models 
	with linear diffusive barriers (1LDB and 2LDB) give results much closer to the correct true 
	values, compared to the original static barriers case  
	2SB \cite{Jennings}. 
	In particular, the 2LDB is within 1-3$\sigma$ concordance for all cases.   

{\large
\begin{table*}
\centering
\caption{Values for best-fit, mean and $1\sigma$ errors in the modified gravity parameters ($f_{R0}$ and $z_{SSB}$) for the three void abundance models 2SB, 1LDB and 2LDB.}
\begin{tabular}{l|c|c|c|c|c|c}
\hline 
Gravity parameters & \multicolumn{3}{c}{Best-Fit} & \multicolumn{3}{|c}{Mean $ \pm$ ($1\sigma$ error)}\\ 
\hline                     
 & 2SB & 1LDB & 2LDB & 2SB & 1LDB & 2LDB \\
\hline
$\log_{10}|f_{R0}| = -8$ ($\Lambda$CDM) & -6.24 & -8.00 & -8.00 & -6.24$\pm$0.09 & -7.94$\pm$0.08 & -7.92$\pm$0.10 \\
$\log_{10}|f_{R0}| = -6$ & -5.78 & -5.88 & -6.04 & -5.79$\pm$0.07 & -5.89$\pm$0.15 & -6.04$\pm$0.14 \\
$\log_{10}|f_{R0}| = -5$ & -5.51 & -4.95 & -5.10 & -5.51$\pm$0.07 & -4.95$\pm$0.16 & -5.09$\pm$0.19 \\
$\log_{10}|f_{R0}| = -4$ & -5.36 & -4.01 & -4.00 & -5.36$\pm$0.08 & -4.09$\pm$0.11 & -4.16$\pm$0.20 \\
\hline     
$z_{SSB} = 0$  ($\Lambda$CDM) & 1.14 & 0.32 & 0.21 & 1.14$\pm$0.04 & 0.27$\pm$0.19 & 0.20$\pm$0.16 \\
$z_{SSB} = 1$ & 1.46 & 1.17 & 1.16 & 1.46$\pm$0.03 & 1.17$\pm$0.05 & 1.17$\pm$0.06 \\
$z_{SSB} = 2$ & 1.63 & 1.89 & 1.88 & 2.31$\pm$0.03 & 1.89$\pm$0.07 & 1.87$\pm$0.07 \\
$z_{SSB} = 3$ & 1.77 & 3.00 & 2.81 & 2.59$\pm$0.03 & 2.97$\pm$0.05 & 2.81$\pm$0.08 \\
\hline   
\end{tabular}
\label{Best_and_mean}
\end{table*}
}

\subsection{Voids in Galaxy Samples}

	In real observations it is much harder to have direct access to the the dark matter density field. 
	Instead we observe the galaxy field, a biased tracer of the dark matter. Therefore it is
	important to investigate the abundance of voids defined by galaxies and the possibility of 
	constraining cosmology and modified gravity in this case.
	
	We introduce galaxies in the original dark matter simulations using the 
	Halo Occupation Distribution (HOD) model from \cite{Zheng}. In \cite{Nadathur2} the authors 
	investigated 
	similar void properties but did not considered spherical voids, using instead the 
	direct outputs of the \texttt{VIDE} \cite{VIDE} void finder.
	
	In our implementation, first we find the dark matter halos in the simulations using the 
	overdensities outputted by 
	\texttt{ZOBOV}. We grow a sphere around each of the densest particles until its enclosed 
	density is $200$ times the mean density of the simulation. This process is the
	reverse analog of the spherical void finder described in \S~IV, the only difference being 
	the criterium used to sort the list of potential halo centers. Here we sort them using the value of 
	the point density, not a S/N significance, as the latter is not provided by 
	\texttt{ZOBOV} in the case of halos.  
	
	We populate these halos with galaxies using the HOD model of \cite{Zheng}. 
	This model consist of a mean occupation function of central galaxies given by
\begin{equation}
\langle N_{cen}(M)\rangle = \frac{1}{2} \left[ 1 + {\rm erf}\left( \frac{\log M - \log M_{min}}{\sigma _{\log M}}\right) \right], 
\end{equation} 
with a nearest-integer distribution. The satellite galaxies follow a Poisson distribution with mean given by
\begin{equation}
\langle N_{sat}(M)\rangle = \langle N_{cen}(M)\rangle \left( \frac{M-M_{0}}{M_{1}'}\right) ^{\alpha}.
\end{equation} 

	Central galaxies are put in the center of halo, and the satellite galaxies are distributed following a 
	Navarro Frenk and White \cite[(NFW),][]{NFW} profile.

	We use parameter values representing the sample {\it Main 1} of \cite{Nadathur2}, namely: 
	$(\log M_{min}, \sigma _{\log M}, \log M_{0}, \log M_{1}', \alpha) = (12.14, 0.17, 11.62, 13.43, 1.15)$. 
	These parameters give a mock galaxy catalogue with galaxy bias $b_g = 1.3$ and mean 
	galaxy density $\bar{n}_g = 5.55 \times 10^{-3} (h/$Mpc$)^{3}$ in $\Lambda$CDM.
	
	We then find voids in this galaxy catalogue using the same algorithm applied to the dark matter 
	catalogue (described in \S~IV). We use the same criterium that a void is a spherical, 
	non-overlapping structure with overdensity equal to $0.2$ times the background 
	galaxy density. However, as the galaxies are a biased 
	tracer of the dark matter field, if we find galaxy voids with $0.2$ times the mean density, we are 
	really finding regions which are denser in the dark matter field. In fact, if 
	$\delta _{g}  = b_{g} \delta$ is the galaxy overdensity, with galaxy bias $b_{g}$ and $\delta$ is the dark matter overdensity 
	we have
\begin{equation}
\Delta = 1+\delta=1+\frac{\delta _{g}}{b_{g}},
\end{equation}
 
	Therefore, if we find voids with $\delta _{g} = -0.8$ and $b_{g} = 1.3$ we have $\Delta = 0.38$,  
	i.e. the galaxy voids enclose a region of density $0.38$ times the mean density of the dark matter field.  
	Therefore it is this value that must be used in the previous theoretical predictions.
	
\begin{figure}
\begin{center}
\includegraphics[width=3.2in]{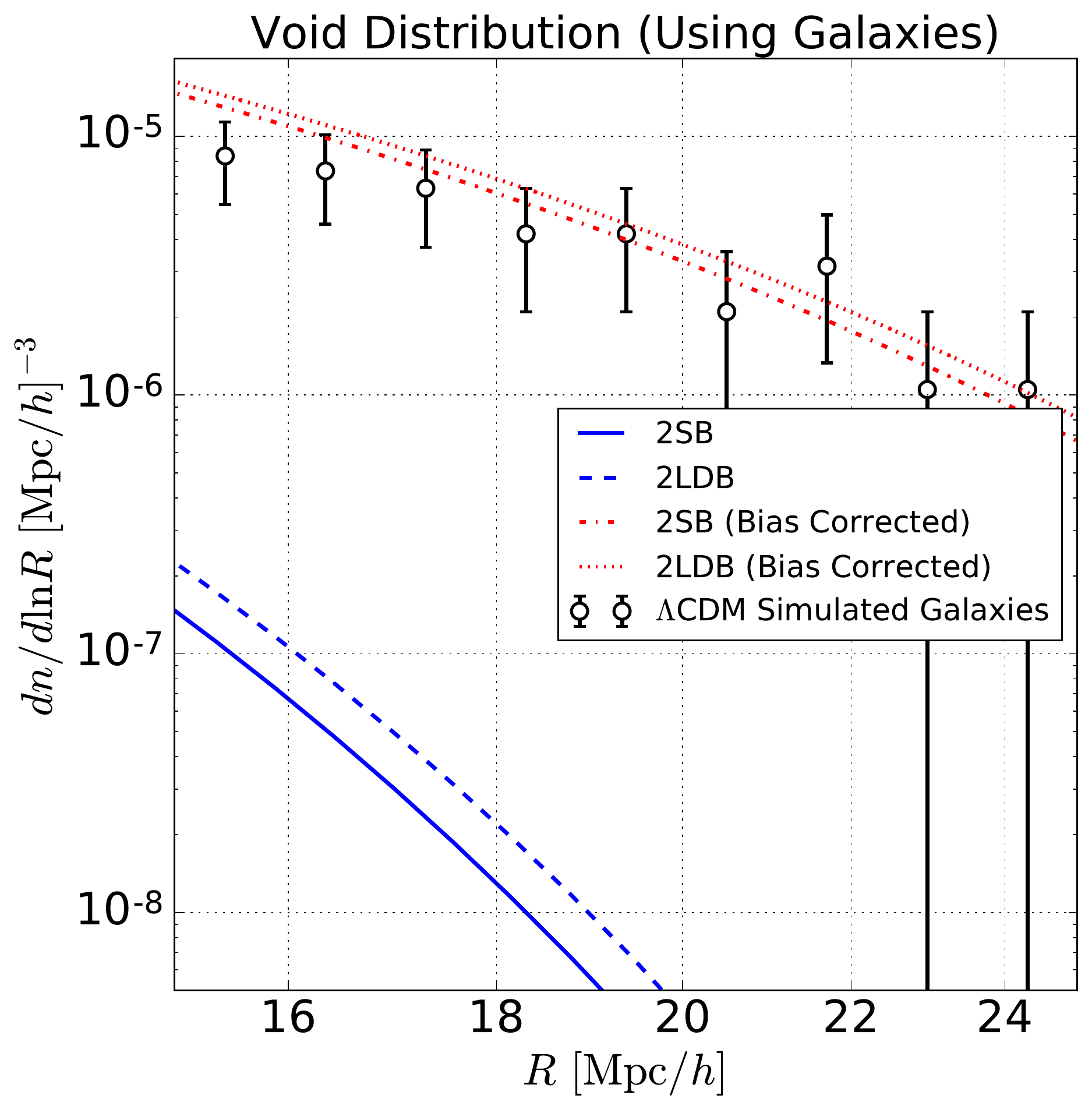}
\caption{Void abundance distribution as a function of void radius 
for voids detected in the {\it galaxy} mock catalogue for $\Lambda$CDM 
(open circles). Also shown are the abundance predictions from the 2SB and 2LDB models 
with no corrections due to galaxy bias (blue solid and dashed lines respectively), 
as well as the same model predictions with 
the bias corrections (red dotted dashes and dotted lines respectively).}
\label{Abundance_gals}
\end{center}
\end{figure}

	Using this value, the relation between linear and nonlinear radii is $R = 1.37 R_{L}$,
	 and the density parameter for the spherical void formation -- calculated using the spherical
	 expansion equations (\S~II.D) -- is $\delta _{v} = -1.33$. We insert these new values 
	 into the theoretical predictions and compare to the measured galaxy void abundance. 
	 The result is shown in Fig.~\ref{Abundance_gals} for the $\Lambda$CDM case. 
	 We see that both original models, 2SB and 2LDB (blue curves), with $R=1.71R_{L}$ and 
	 $\delta _{v} = -2.788$, provide incorrect predictions for the abundance of galaxy voids. 
	 However when corrected for the galaxy bias (red curves), these models are in good agreement 
	 with the data. We also see that the 2LDB provides a slightly better fit, which is not 
	 significant given the error bars.

\begin{figure}
\begin{center}
\includegraphics[width=3.2in]{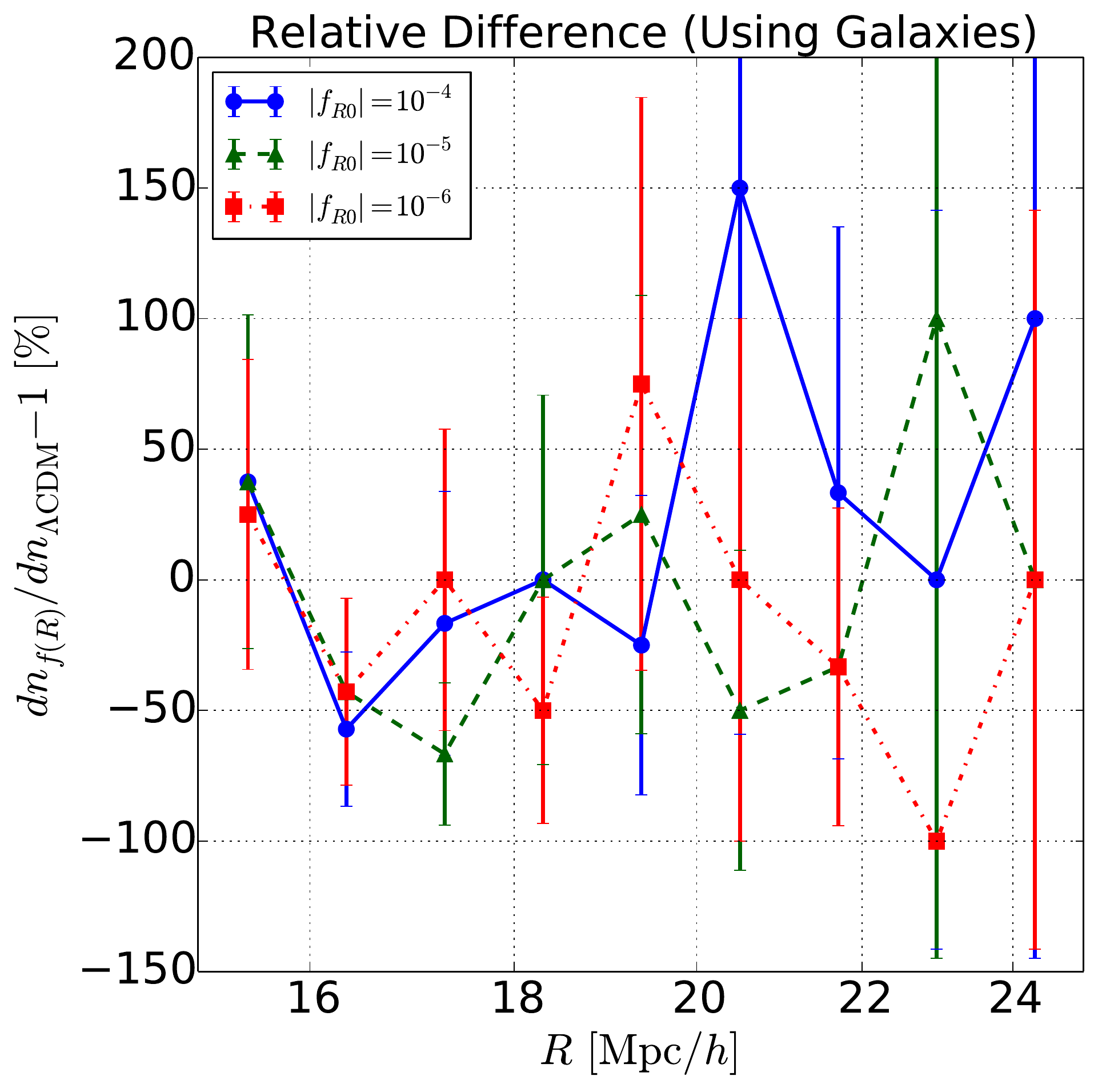}
\caption{Relative difference in galaxy void abundance as measured in $f(R)$ gravity simulations and 
GR simulations. The difference is shown for $|f_{R0}| = 10^{-6}$ (red squares), $10^{-5}$ (green triangles) and $10^{-6}$ (blue circles).
}
\label{Diff_gals}
\end{center}
\end{figure}

	The main problem of our galaxy catalogues is the low number density of objects. 
	 Larger box sizes (or a galaxy population intrinsically denser) 
	 might help decrease the error bars sufficiently in order to constrain modified 
	 gravity parameters. 
	 In Fig.~\ref{Diff_gals} we show the relative difference between the abundance for the 
	 three modified gravity models and GR as inferred from our simulations. 
	 We see that it is not possible to constrain the gravity model using the abundance of 
	 galaxy voids, as extracted from mock galaxy catalogues of the size considered here, 
	 due to limited statistics. 
	 Further investigations using larger or multiple boxes, or else considering a 
	 galaxy population with larger intrinsic number density should decrease Poisson errors 
	 significantly, allowing for a better investigation of void abundance in the large  
	 data sets expected for current and upcoming surveys, such as the SDSS-IV, DES, DESI, 
	 Euclid and LSST.  
	
\section{Discussion and Conclusion} \label{sec:discussion}

We have used a suite of N-body simulations from the Isis code \cite{Llinares} for GR and modified 
gravity models to define spherical voids from underdensities detected by {\tt ZOBOV} \cite{Neyrinck}, 
a void-finder based on Voronoi tesselation.  We find that the void abundance in modified gravity and 
$\Lambda$CDM may differ by $\sim 100\%$ for the largest void radii in our simulations.

We interpreted the void abundance results through a spherical expansion model and extended 
Excursion Set approach. The most general theoretical model considered has two drifting diffusive 
barriers, with a linear  dependence on the density variance (2LDB, see \S~III). This model depends 
on the theory linear power spectrum $P(k)$ and in principle has multiple parameters, namely 
$\delta_c$ and $\delta_v$ (the critical densities for collapse and expansion),  $\beta_c$ and 
$\beta_v$ (the barrier slopes for halos and voids) and $D_c$ and $D_v$ (the diffusion coefficients 
for halos and voids). Fixing $\delta_c$ and $\delta_v$ to their GR values and under the simplifying 
assumption that $\beta=\beta_c=\beta_v$, the model depends on two free parameters: $\beta$ and 
$D=D_c+D_v$. 
Interestingly, our model accounts for the void-in-cloud effect and 
generalizes previous models for void abundance based on static barriers \cite{Jennings}. 
The generalizations proposed here are similar to those made by \cite{Maggiore1, Maggiore2} in the 
context of halos.

Since our model requires the linear power spectrum in modified gravity, we have implemented a 
numerical evolution of the linear perturbation equations for general theories of modified gravity 
parametrized by Eq.~\eqref{delta_lin}. We compared our computation to that from {\tt MGCAMB}
 for $f(R)$ gravity and found very good agreement. We then use this implementation to compute 
 the linear spectrum for both $f(R)$ and symmetron gravity.

We also considered approximate equations for spherical collapse and spherical expansion and derived the 
spherical collapse parameters $\delta_c$ and $\delta_v$ as a function of scale, recovering in particular 
the values in the strong and weak field regimes of $f(R)$ gravity -- the latter corresponding to the GR solution. 
We then estimated the dependence of barriers $B_c$ and $B_v$ with the variance $S$ 
and derived values for $\beta_{c,v}$ and $\delta_{c,v}$. The values found did not however 
seem to correctly describe the void abundance from simulations, which may be due to the 
approximated equations used to study the expansion/collapse.

We also found that the variations on $P(k)$, $\beta$ 
and $D$ as a function of modified gravity were much stronger than those from $\delta_c$ and $\delta_v$.   
Therefore, in our modeling of void abundance we kept $\delta_c$ and $\delta_v$ fixed to 
their GR values, and took $\beta$ and $D$ as free parameters to be fit from simulations. 
 Although beyond the scope of this work, we envision that it should be possible to derive 
the model parameters from first principles in the future. 

By comparing the measured void abundance from the simulations to the theoretical models considered, 
we found the best fit values for $\beta$ and $D$ in each gravity theory and each abundance model. 
In particular, we found that these parameters were best-fit for 
models with linear diffusive barriers (see Figs.~\ref{Distribution_fofr},~\ref{Distribution_symm} 
and Table~\ref{chi2}), indicating that the addition of these features is important 
to describe modified gravity effects on void abundance. 
This allowed us to then fit for $\beta$ and $D$ as a function of modified gravity parameters, namely 
$|f_{R0}|$ in the case of $f(R)$ gravity, and $z_{SSB}$ in the case of symmetron. 

Next we used these fits to check how well the calibrated models could recover the modified gravity 
parameters from hypothetical and idealized void abundance observations. We compared the void 
abundance measured in simulations to the model predictions and performed an MCMC search 
for the gravity parameters. Since the predictions were calibrated from the simulations themselves, our 
results may be highly optimistic. Nonetheless, we found that the models 
with linear diffusive barriers recover the modified gravity parameters better than the model 
with static barriers for all gravity theories (see Figs.~\ref{Posterior_fofr},~\ref{Posterior_symm} 
and Table~\ref{Best_and_mean}). 
We also found that when using voids found in the GR simulation to fit for modified gravity
parameters, we seem to properly recover the GR limit at the $2\sigma$ level. Since we only used one simulation for each 
gravity model considered, our results have considerable uncertainties. We expect these 
to improve significantly with the use of multiple and larger simulations. 

Finally, we populated the dark matter halos found in the simulations with galaxies in order to 
access the possibility of modeling the abundance of galaxy voids. For the GR case, we found 
that the same model with linear diffusive barriers properly describes the abundance of 
galaxy voids, provided we use the galaxy bias to correct for the effective overdensity $\Delta$ 
used for void detection. However, the error bars were too large to allow for any signal in the 
modified gravity case relative to GR. Again since we  
used a single simulation for each gravity, our results for galaxy voids are even more affected by 
shot noise and unknown sample variance effects. 

Current and upcoming spectroscopic and photometric galaxy surveys will produce
large catalogs of galaxies, clusters and voids.
Observed void properties from real data are affected by nontrivial effects such as 
surveys masks and depth variations in the sky. One could partially characterize these effects 
from realistic simulations and understand their possible consequences, such as 
inappropriately breaking large voids into multiple smaller ones or vice-versa 
(i.e. merging small voids into larger ones). 
Assuming that such effects can be understood and characterized, we expect that 
the properties of voids, including their abundance, clustering properties and profiles, will be very important 
to constrain cosmological models, especially modified gravity. In particular, since voids 
and halos respond differently to screening effects present in viable modified gravity 
theories, a combination of voids and halo properties should be 
particularly effective in constraining and distinguishing alternative gravity models.

\section*{Acknowledgments}
RV is supported by FAPESP. 
ML is partially supported by FAPESP and CNPq.
CLL acknowledges support from the STFC consolidated grant ST/L00075X/1.
DFM acknowledges support from the Research Council of Norway, and the NOTUR facilities. 

\bibliography{Void_distribution}

\end{document}